\documentclass[a4paper,11pt]{article}
\pdfoutput=1
\usepackage{jheppub} 
\usepackage{graphicx}
\usepackage{amsmath}
\usepackage{amssymb}
\usepackage{bbold}
\usepackage{float}
\usepackage{slashed}
\usepackage{dcolumn}
\usepackage{bm}
\usepackage{color}
\usepackage{multirow}
\allowdisplaybreaks

\usepackage{tabularx}
\usepackage{makecell}

\newcommand{\lsim}{{\;\raise0.3ex\hbox{$<$\kern-0.75em\raise-1.1ex\hbox{$\sim$}}\;}}
\newcommand{\gsim}{{\;\raise0.3ex\hbox{$>$\kern-0.75em\raise-1.1ex\hbox{$\sim$}}\;}}
\def\bea{\begin{eqnarray}}
\def\eea{\end{eqnarray}}
\def\bec{\begin{center}}
\def\ec{\end{center}}

\def\beq{\begin{equation}}
\def\eeq{\end{equation}}

\def\bea{\begin{eqnarray}}
\def\eea{\end{eqnarray}}
\def\beq#1\eeq{\begin{align}#1\end{align}}
\def\beqnn#1\eeq{\begin{align*}#1\end{align*}}
\def\ba{\begin{array}}
\def\ea{\end{array}}
\def\bc{\begin{center}}
\def\ec{\end{center}}

\newcommand{\dis}[1]{\begin{equation}\begin{split}#1\end{split}\end{equation}}

\def\gev{\rm GeV}
\def\tev{\rm TeV}
\def\mev{{\rm MeV}}

\usepackage{breqn} 
\makeatletter 
\let\cref@old@eq@setnumber\eq@setnumber 
\def\eq@setnumber{% 
\cref@old@eq@setnumber% 
\cref@constructprefix{equation}{\cref@result}% 
\protected@xdef\cref@currentlabel{% 
[equation][\arabic{equation}][\cref@result]\p@equation\theequation}} 
\makeatother

\usepackage{cases}

\newcommand{\be}{\begin{equation}\begin{aligned}}
\newcommand{\ee}{\end{aligned}\end{equation}}

\newcommand{\beqa}{\begin{eqnarray}}
\newcommand{\eeqa}{\end{eqnarray}}

\renewcommand{\eqref}[1]{Eq.~(\ref{#1})}

\newcommand{\eg}{{\em e.g.}}
\newcommand{\ie}{{\em i.e.}}

\newcommand{\TODO}[1]{\textcolor{green}{TODO}}
\RequirePackage[normalem]{ulem}

\DeclareUnicodeCharacter{2212}{\textendash}

\makeatletter
\def\l@subsubsection#1#2{}
\makeatother

\preprint{CTPU-PTC-24-41}

\title{Searches for Power-Law Warped Extra Dimensions}

\author{
    Sang Hui Im\footnote{email: imsanghui@ibs.re.kr} and
    Krzysztof Jod\l{}owski\footnote{email: k.jodlowski@ibs.re.kr}
 }
     
\affiliation{
    Particle Theory and Cosmology Group, Center for Theoretical Physics of the Universe, \\
    Institute for Basic Science (IBS), Daejeon 34126, Korea \
}

\abstract{
    Extra dimensions with a bulk dilaton field can be power-law warped, unlike the exponential warping in the Randall-Sundrum (RS) model. We show that this mildly warped extra dimension can address the hierarchy problem with a novel Kaluza-Klein (KK) spectrum characterized by lighter feebly coupled KK modes compared to the KK modes in the RS model. We investigate the prospects of searching for signatures of such KK modes at current and future colliders, such as the LHC, CLIC, and FCC-ee using visible decays of KK gravitons. We also update the current bounds and projected limits for the RS model and the linear dilaton (LD) model.  Furthermore, we explore the long-lived regime of KK gravitons at beam dump experiments, \eg, FASER2, MATHUSLA, and SHiP, as well as constraints from astrophysical and cosmological observations. We find that combining both kinds of searches will enable comprehensive coverage of the model parameter space relevant to the electroweak hierarchy problem. 
} 

\vspace{4cm}

\begin{document} 
\maketitle
\flushbottom

\section{Introduction}
Extra spatial dimensions beyond the observed 4-dimensional spacetime are an intriguing possibility for physics beyond the Standard Model. 
String theory requires such extra dimensions for the consistency of the theory, see e.g. \cite{Polchinski:1998rq,Tong:2009np}. Large flat extra dimensions proposed by Arkani-Hamed, Dimopoulos, Dvali (ADD)~\cite{Arkani-Hamed:1998jmv} and a small warped extra dimension by Randall and Sundrum (RS) \cite{Randall:1999ee} have been shown to be able to address the hierarchy problem with predictions on the new particles with observable signatures around the TeV scale.
Similarly, a 5D analogue of 7D theory with a linear dilaton background, which provides a dual description of Little String Theories, was proposed to solve the hierarchy problem with distinctive Kaluza-Klein (KK) graviton signatures, which is often referred to as the ``linear dilaton model" \cite{Antoniadis:2011qw, Giudice:2017fmj}. 

Later, an interesting observation was made that the linear dilaton (LD) model can be obtained by taking the continuum limit of the so-called clockwork Lagrangian \cite{Giudice:2016yja}, which was originally introduced to generate hierarchical couplings of a compact field \cite{Choi:2014rja, Choi:2015fiu, Kaplan:2015fuy}. Generalizing this observation, \cite{Choi:2017ncj} showed that an extra dimension with a bulk dilaton field having a general coupling (hence ``\emph{general linear dilaton} (GLD) model") can have more general geometries beyond the ADD model (flat), the RS model (exponentially warped), and the LD model (linearly warped). Using the notation of \cite{Choi:2017ncj}, the metric of this general geometry is given by
\dis{
ds^2 = e^{2 k_1 y} \eta_{\mu \nu} dx^{\mu} dx^{\nu} + e^{2k_2 y} dy^2, \label{metric0}
}
where $\mu, \nu =0, 1, 2, 3$ are 4D spacetime indices, $\eta_{\mu \nu} = \textrm{diag}(-1, 1, 1, 1)$, $y$ is a coordinate for the extra dimension, and $k_1, k_2$ are two independent real parameters (but with the same sign) determined by the dilaton potential. 
By changing the coordinate for the extra dimension from $y$ to $z$ such that $dz = \exp(k_2 y) dy$, this metric is equivalent to
\dis{
ds^2 = (k_2 z+1)^{2k_1/k_2} \eta_{\mu \nu} dx^{\mu} dx^{\nu} + dz^2. \label{metric1}
}
The previous three well-known extra dimensional models are then identified with $k_1 = 0$ (ADD), $k_1=k_2$ (LD), and $k_2 \rightarrow 0$ (RS). We also note that $k_1/k_2 = 1/6, 1/7,$ and $1/10$ are shown to be realized by 5D effective theory of heterotic M-theory \cite{Lukas:1998tt, Im:2018dum}, and $k_1/k_2=(N+2)/(N-1)$ by 5D effective theory of $(4+N)$-dimensional spacetime with $N$ compact extra dimensions having a bulk cosmological constant \cite{Teresi:2018eai}.\footnote{In Ref. \cite{Teresi:2018eai}, the metric is given in the Jordan frame, which is equivalent to the metric (\ref{metric0}) or (\ref{metric1}) in the Einstein frame.}

It was pointed out that LD ($k_1/k_2=1$) has a distinctive feature for the KK graviton spectrum characterized by more degenerate masses compared with RS  \cite{Antoniadis:2011qw, Giudice:2016yja, Giudice:2017fmj}.
As we will explain in the next section, and also argued in \cite{Im:2018dum}, a warped extra dimension with $0< k_1/k_2 < 1$ can have even more distinctive patterns for KK graviton spectrum and couplings in contrast to the conventional extra dimensional models such as RS and ADD. As naively expected, in this new class of extra dimensions, the KK modes have intermediate features between RS and ADD. Here, let us briefly summarize the characteristics of the KK graviton spectrum and the couplings to the Standard Model (SM) particles for each class of models. The 4D effective Lagrangian for the KK gravitons may be written as 
\dis{
{\cal S} = \int d^4 x \left[ -\frac12 (\partial_\rho h_{\mu \nu}^{(0)})^2  -\frac{h_{\mu \nu}^{(0)}}{M_P} T_{\rm SM}^{\mu \nu} -\sum_{n=1}^\infty \left(\frac12 (\partial_\rho h_{\mu \nu}^{(n)})^2 + \frac12 m_n^2 (h_{\mu \nu}^{(n)})^2+  C_n \frac{h_{\mu \nu}^{(n)}}{M_5} T_{\rm SM}^{\mu \nu} \right) \right]
\label{eq:lagr_4D}
}  
where $h_{\mu \nu}^{(0)}$ is the massless zero mode graviton, $h_{\mu \nu}^{(n)}$ with $n \geq 1$ is the KK graviton with mass $m_n$ and coupling $C_n$ to the energy-momentum tensor $T_{\rm SM}^{\mu \nu}$ of the SM fields, $M_P$ is the 4D reduced Planck mass, and $M_5$ is the 5D Planck mass, which is around TeV scale in order to address the electroweak hierarchy problem.\footnote{Here we take the convention that the IR brane is located at $y=0$ with the metric (\ref{metric0}) where the SM particles are localized, while the UV brane is located at $y=\pi R$. Then $M_5$ is related to $M_P$ by the relation: $M_P^2 \simeq M_5^{3} e^{2k \pi R}/k$ where $k=k_1+k_2/2$.} For each class of models, it can be shown that
\bea
\textrm{RS : } &&  m_n \sim n k, \quad C_n \sim {\cal O}(1) \nonumber \\
\textrm{LD : } &&  m_n \sim k \sqrt{1+ \left(\frac{n\pi}{\ln (M_P/M_5)} \right)^2}, \quad C_n \sim \sqrt{\frac{k}{M_5\ln (M_P/M_5)}}  \left[1+ \left(\frac{\ln (M_P/M_5)}{n\pi} \right)^2 \right]^{-\frac12} \nonumber\\
\textrm{ADD :} && 
m_n \sim n M_5 \left(\frac{M_{5}}{M_P} \right)^2, \quad C_n \sim \frac{M_5}{M_P} \nonumber \\
\textrm{GLD :} &&  m_n \sim n k \left(\frac{M_5}{M_P} \sqrt{\frac{M_5}{k}} \right)^{\frac{2(1-k_1/k_2)}{(1+2k_1/k_2)}} , \quad C_n \sim \frac{M_5}{M_P} n^{\frac{3k_1/k_2}{2(1-k_1/k_2)}}
\eea 
where $k \,(\lesssim M_5)$ corresponds to the curvature scale of the extra dimension, which can be naturally smaller than $M_5$ in LD and GLD due to dilaton shift symmetry \cite{Giudice:2016yja, Giudice:2017fmj}, and we take $0<k_1/k_2<1$ for GLD here.
In the case of RS, the KK spectrum and couplings are mostly determined by the scale $ M_5\, (\sim k)$. It is similar in the case of LD if the curvature $k$ is comparable to $M_5$, except for the feature that the KK spectrum is more compressed. On the contrary, the KK modes of ADD are much lighter than $M_5$, and their couplings are as small as the zero mode graviton coupling, which is  suppressed by the 4D Planck scale. On the other hand, the KK gravitons of a power-law warped extra dimension with $0 < k_1/k_2 < 1$ can be parametrically lighter than $M_5$, similar to ADD, while having sizable couplings larger than the zero mode graviton coupling for large $n$. 

The pattern of the KK spectrum and couplings of GLD with $0 < k_1/k_2 < 1$ is thus similar to ADD for lightest KK states, while it shows substantial difference for heavy KK states with large $n$ characterized by sizable couplings that may allow larger production of the heavy KK states at colliders.
The ADD model with one extra dimension is heavily constrained by astrophysical processes, in particular the Supernova 1987A (SN1987), which require $M_5\gtrsim 700 \,\tev$ \cite{Hannestad:2003yd}, and therefore cannot provide a \textit{natural} solution to the hierarchy problem.
On the other hand, the phenomenology of GLD with $0<k_1/k_2<1$ has not been comprehensively studied before, while it can be realized in some concrete UV setups, e.g. heterotic string theory \cite{Lukas:1998tt, Im:2018dum}. In this work, we found that GLD with $k_1/k_2  \gtrsim 1/3$ can provide a solution to the hierarchy problem with the natural Higgs mass scale $\sim M_5$ less than 10 TeV with some predictions on KK graviton signatures in future colliders and relatively weak constraints from astrophysical objects, while GLD with $0<k_1 / k_2 \lesssim 1/3$ is rather strongly constrained by astrophysical bounds on $M_5$. 

Another interesting point for LD and GLD phenomenology is that the curvature $k$ of the extra dimension can be parametrically smaller than $M_5$ thanks to the dilaton shift symmetry, as discussed in \cite{Giudice:2016yja, Giudice:2017fmj}.  
The curvature $k$ breaks the translation invariance, allowing decays of heavier gravitons to lighter ones, which do not happen in ADD, because the decay would happen at threshold~\cite{Arkani-Hamed:1998jmv}, where phase space vanishes. The branching ratios are small if $k$ is comparable to $M_5$ which will be natural in the case of RS where there is no symmetry for $k \rightarrow 0$. On the other hand, if $k$ is small enough, the branching ratios can be sizable. 
As a result, in the low curvature region of LD, $k\lesssim 10\,\gev$, the SM signatures at the LHC are suppressed because the heavier KK states decay into lighter ones with significant branching ratios, while the lighter states are long-lived and escape the LHC detectors \cite{Giudice:2016yja,ATLAS:2017fih,CMS:2017yta}.
On the other hand, such long-lived regime can be covered by present and future electron or proton beam dump experiments, see \cite{Lanfranchi:2020crw} for a recent review, which look for, \eg, $\sim\gev$ mediators between the SM and the dark sector such as axion-like particles (ALPs), dark photon, dark Higgs, or sterile neutrinos, among others. We will show that in addition to collider searches, beam dumps can also be effective in constraining extra dimensional theories by looking for visible decays of KK states. 

This paper is organized as follows. In section \ref{sec:theory}, we discuss power-law warped extra dimensions originating from GLD, in particular the spectrum and couplings of the KK gravitons, which are crucial for phenomenology. In section \ref{sec:searches}, we discuss the signatures of the KK gravitons from RS, LD, and GLD at the LHC, upcoming lepton colliders, beam dumps, and astrophysical processes. In section \ref{sec:EAlimits}, we present and discuss our results - the main ones being sensitivity plots of these experiments and observations. We also discuss the importance of KK graviton decays into lighter KK states. In particular, we find the LHC searches for the LD are stronger than the previous results in Ref. \cite{Giudice:2017fmj} in the $k\ll M_5$ region.
We conclude in section \ref{sec:conclusions}, while auxiliary sections are in the appendix. 
In appendix \ref{app:GLDBg}, we present the  derivation of the GLD solution.
In appendix \ref{app:astro_bounds_ADD}, we succinctly recall the astrophysical bounds on flat large extra dimensions, and we present the revised (weaker) limits.
Technical details of our analysis are relegated to appendix \ref{app:xs}.

\section{Power-law warped extra dimensions} \label{sec:theory}
In this section, we will show that an extra dimension with a general warping in Eq. (\ref{metric1}) can be realized by a bulk dilaton field background. Subsequently, we will discuss the spectrum and couplings of the KK gravitons relevant for our phenomenological analysis. The discussion will closely follow Ref. \cite{Choi:2017ncj, Im:2018dum}. 

\subsection{General Linear Dilaton model}
The GLD model introduced in \cite{Choi:2017ncj} may be defined as the following 5D action in the Jordan frame:
\dis{
{\cal S}=\int d^4x \int_{-\pi R}^{\pi R} dy \sqrt{-g}\, e^{\xi S} &\left( \frac{M_5^3}{2} R_5 + \frac{M_5^3}{2} \partial^M S \partial_M S - \Lambda_5 
-\frac{1}{\sqrt{g_{55}}} \left[\Lambda_0 \delta(y) + \Lambda_\pi \delta(y-\pi R) \right]\right) \,,
\label{GLD_J}
}
where the extra dimension is described by the coordinate $y \in [-\pi R, \pi R]$ and orbifolded with $Z_2$-identification $y\equiv -y$, $M_5$ is the 5D Planck mass, $\Lambda_5$ is the bulk constant parameter, $\Lambda_0$ and $\Lambda_\pi$ are the tensions of 3-branes localized at $y=0$ and $y=\pi R$, respectively, and $R_5$ is the 5D Ricci scalar. $S$ is the dilaton field propagating over the extra dimension, and interactions involving the dilaton field are determined by the overall factor $e^{\xi S}$  in the Jordan frame with a real coupling constant $\xi$. This form of dilaton interactions may be motivated by the \emph{classical scale invariance} of effective actions originating from string theory \cite{Witten:1985xb,Dine:1985kv}. It requires that classical equations of motion of the action ${\cal S}$ do not change under the scale transformation realized by dilaton shift symmetry $S \rightarrow S + \alpha$ in the Jordan frame. For further discussion of this point, we refer the readers to section 2.2 of \cite{Giudice:2017fmj}.

The action (\ref{GLD_J}) can be written in the Einstein frame by redefining the metric as
\dis{
g_{MN} \rightarrow e^{-2\xi S/3} g_{MN},
}
where Latin alphabets stand for 5D spacetime indices $M, N =0, 1, 2, 3, 5$.  
Furthermore, setting the canonical normalization of the dilaton field, the action becomes
\dis{
{\cal S}=\int d^4x \int_{-\pi R}^{\pi R} dy \sqrt{-g}\, &\left( \frac{M_5^3}{2} R_5 - \frac{M_5^3}{2} \partial^M S \partial_M S - e^{-2\lambda S/\sqrt{3}} \Lambda_5 
\right. \\
&\quad \left. -\frac{e^{-\lambda S/\sqrt{3}}}{\sqrt{g_{55}}} \left[\Lambda_0 \delta(y) + \Lambda_\pi \delta(y-\pi R) \right]\right)\,, \label{GLD_E}
}
where the dilaton coupling $\lambda$ in the Einstein frame is given by 
\dis{
\lambda = \frac{\xi}{\sqrt{4\xi^2-3}}.
}
Here we note that it is needed that
\dis{
\xi^2 > \frac34 \label{ghost}
}
in order for the dilaton $S$ not to be a ghost field in the Einstein frame. It will turn out to be convenient to use parametrization 
\dis{
\Lambda_5 = -2M_5^3 k_b^2, \quad \Lambda_0 = -4 M_5^3 k_0, \quad \Lambda_\pi = -4M_5^3 k_\pi.
}
As discussed in \cite{Giudice:2016yja, Giudice:2017fmj}, we remark that the shift symmetry $S \rightarrow S + \alpha$ in the Einstein frame is restored in the bulk as $k_b\rightarrow 0$. Therefore, small $k_b \ll M_5$ is technically natural in GLD models - contrary to RS, where the dilaton is decoupled - and we will examine its phenomenological consequences in subsequent sections.   

A solution to the equations of motion for the metric field $g_{MN}$ and the dilaton field $S$ of the action (\ref{GLD_E}) is given in \cite{Choi:2017ncj}; it is also derived in appendix \ref{app:GLDBg}. It is 
\dis{
ds^2 = e^{2k_1 |y|} (\eta_{\mu \nu} dx^\mu dx^\nu) + e^{2k_2 |y|} dy^2, \quad \frac{\lambda}{\sqrt{3}}S = k_2 |y|+\frac{\lambda}{\sqrt{3}}S_0\,,
\label{GLD_sol}
}
where Greek alphabets stand for 4D spacetime indices $\mu, \nu =0, 1, 2, 3$, 
\dis{
k_1 =  \frac{2k_b}{\sqrt{3(4-\lambda^2)}} e^{-\lambda S_0/\sqrt{3}}, \quad k_2 =\lambda^2 k_1\,,
}
and $S_0$ is the vacuum expectation value of $S$ at $y=0$ which can be determined by an additional brane potential for $S$ as demonstrated in \cite{Giudice:2017fmj}.
Also the brane tension terms $\Lambda_0$ and $\Lambda_\pi$ have to satisfy
\dis{
k_0 e^{-\lambda S_0/\sqrt{3}}= -k_\pi e^{-\lambda S_0/\sqrt{3}} = \frac32 k_1 \label{k0pi}
}
in order to have the 4D Minkowski background. 
In Eq. (\ref{GLD_sol}), the dilaton field solution is linear in the extra dimensional coordinate, i.e. $S \propto y$. This is the reason why this model is often called \emph{linear dilaton} model. 
Also, it is \emph{general} linear dilaton model, since the general dilaton coupling $\xi$ is introduced, while $\xi=1$ in the original linear dilaton model studied in \cite{Antoniadis:2011qw, Giudice:2017fmj}.

Therefore, the metric solution in Eq. (\ref{GLD_sol}) with the linear dilaton background realizes a generally warped extra dimension, for which the metric can be rewritten as Eq. (\ref{metric1}) by changing the extra dimensional coordinate from $y$ to $z$ such that $dz = e^{k_2 |y|} d|y|$:
\dis{
ds^2 = (k_2 z+1)^{2k_1/k_2} \eta_{\mu \nu} dx^\mu dx^\nu + dz^2\,, \label{metric1'}
}
where $z \in [0, (e^{k_2 \pi R}-1)/k_2]$ and
\bea
\frac{k_1}{k_2} &=& 4-\frac{3}{\xi^2}\,, \label{k1/k2} \\
k_2 &=& \frac{2 \xi^2}{3(4\xi^2-3)}\sqrt{\frac{4\xi^2-3}{5\xi^2-4}} k_b e^{-\lambda S_0/\sqrt{3}}\,. \label{k2}
\eea
Thus, the dilaton coupling $\xi$ in the Jordan frame action (\ref{GLD_J}) determines the power $(=k_1/k_2)$ of the warping, while the bulk constant parameter $\Lambda_5=-2M_5^3 k_b^2$ and the dilaton vacuum value $S_0$ at $y=0$ controls the curvature factor $k_2$. Moreover, Eq. (\ref{k1/k2}) tells us that $k_1/k_2>0$ from the ghost-free condition (\ref{ghost}), which means that $k_1$ and $k_2$ must have the same sign.

\subsection{Spectrum and couplings of Kaluza-Klein gravitons}
The KK gravitons arise from the  tensor fluctuation $h_{\mu \nu}(x, y)$ of the 4D part of the metric on the background given by Eq. (\ref{GLD_sol}):
\dis{
ds^2 = e^{2k_1 |y|} (\eta_{\mu \nu} +2h_{\mu \nu} (x, y)) dx^\mu dx^\nu + e^{2k_2 |y|} dy^2.
}
The effective action for the 4D metric fluctuation can be found by inserting the above metric to the action (\ref{GLD_E}). Working in the traceless-transverse gauge, $h^\mu_\mu = \partial^\mu h_{\mu \nu} = 0$, the action for the fluctuation to quadratic order in $h_{\mu \nu}$ is
\dis{
{\cal S}_{h} = -M_5^3 \int d^4x \int_{-\pi R}^{\pi R}  dy \,e^{(2k_1 + k_2) |y|} \left[ \frac12 (\partial_\rho h_{\mu \nu})^2 + \frac12 e^{2(k_1-k_2) |y|} (\partial_y h_{\mu \nu})^2 \right]. \label{Sh}
} 
The equation of motion for the graviton field $h_{\mu \nu}(x, y)$ from the above action comes out as
\dis{
\partial_y^2 h_{\mu \nu} + 2(k + p) \epsilon(y) \partial_y h_{\mu \nu} + e^{-2 p |y|} \eta^{\rho \sigma} \partial_\rho \partial_\sigma h_{\mu \nu} = 0\,, \label{eom}
}
where $\epsilon(y) \equiv \partial_y |y|$, and 
\dis{
k &\equiv k_1 + \frac{k_2}{2}\,, \\
p &\equiv k_1 - k_2\,. \label{def_kp}
}
To find a solution to Eq. (\ref{eom}), let us decompose $h_{\mu \nu}(x, y)$ as
\dis{
h_{\mu \nu}(x, y) = \sum_{n=0}^{\infty}  h_{\mu \nu}^{(n)} (x) h_n(y)\,,
}
where we assume that  $h_{\mu \nu}^{(n)} (x)$ is a 4D mass eigenstate with mass $m_n$, 
\dis{
 \eta^{\rho \sigma} \partial_\rho \partial_\sigma h_{\mu \nu}^{(n)}(x) = m_n^2   h_{\mu \nu}^{(n)}(x). \label{4Deom}
}
Using Eq. (\ref{4Deom}), the equation of motion (\ref{eom}) implies
\dis{
\partial_y^2 h_n(y) + 2(k + p) \epsilon(y) \partial_y h_n(y) + e^{-2 p |y|} m_n^2 h_n(y) = 0.
\label{heq}
}

The solutions to Eq. (\ref{heq}) for $p \neq 0$ turn out to be given by Bessel functions:
\bea
h_0(y) &=& \frac{1}{N_0} ~\textrm{with}~ m_0=0, \\
h_n(y) &=& \frac{1}{N_n} e^{-(k+ p)|y|} \left[ J_\alpha \left( \frac{m_n}{|p|} e^{-p|y|} \right) + c_n Y_\alpha \left( \frac{m_n}{|p|} e^{-p|y|} \right)\right] ~\textrm{with}~ m_n \neq 0,
\eea
where $N_n$ is a normalization factor, $\alpha =  1+ k/p$, $J_\alpha$ and $Y_\alpha$ are the Bessel functions of the first kind and the second kind, respectively. The coefficient $c_n$ is determined by the following boundary values of the Bessel functions, in order to satisfy Eq. (\ref{heq}):
\dis{
c_n =   - \frac{J_{\alpha -1} \left(\frac{m_n}{|p|} \right)}{Y_{\alpha -1} \left(\frac{m_n}{|p|} \right)} = - \frac{J_{\alpha -1} \left(\frac{m_n}{|p|} e^{-p\pi R} \right)}{Y_{\alpha -1} \left(\frac{m_n}{|p|} e^{-p \pi R} \right)}\,. \label{cn}
}
For the special case $p=0$, the solutions to Eq. (\ref{heq}) are given by
\bea
h_0(y) &=& \frac{1}{N_0} ~\textrm{with}~ m_0=0, \\
h_n(y) &=& \frac{1}{N_n} e^{-k|y|} \left[ k \sin\left(\frac{n}{R} |y| \right) + \frac{n}{R} \cos\left(\frac{n}{R} |y| \right) \right] ~\textrm{with}~ m_n \neq 0, \label{hnp0}
\eea
where $n$ is a positive integer. This special case corresponds to the original linear dilaton model given in \cite{Antoniadis:2011qw, Giudice:2017fmj}, which is defined by $\xi=1$ in Eq. (\ref{GLD_J}). 

Eq. (\ref{cn}) and the solution (\ref{hnp0}) determine the mass eigenvalue $m_n$ of the KK graviton mode $h_{\mu \nu}^{(n)} (x)$ as
\dis{
m_n 
\begin{cases}
\simeq \left(n-\frac14 + \left|\frac{k}{2p}\right|\right) \pi p\,,&  p>0 \\
 = \sqrt{k^2 + \frac{n^2}{R^2}}\,, & p=0\\
 \simeq \left(n-\frac14 + \left|\frac{k}{2p}\right|\right) \pi |p| e^{-|p| \pi R}\,,&  p<0  
 \end{cases} \label{KKm}
}
where $n > 1/4 - |k/2p|$ is an integer for $p\neq0$ and a positive integer for $p=0$.

Let us now discuss couplings of the KK gravitons to the SM particles. 
We assume that the SM particles are localized on the brane located at $y=0$, and we take $k_2= \lambda^2 k_1 >0$ in order to address the hierarchy problem.
The KK gravitons couple to the brane-localized SM Lagrangian ${\cal L}_{\rm SM}$ by the interaction
\dis{
{\cal S}_{\rm int} = \int d^4x \int_{-\pi R}^{\pi R} dy \sqrt{-g} \,\frac{1}{\sqrt{g_{55}}} \delta(y)  {\cal L}_{\rm SM} \,,
}
which gives rise to
\dis{
{\cal S}_{\rm int} = \int d^4 x \left[ {\cal L}_{\rm SM} -h_{\mu \nu} (x, 0) T^{\mu \nu}_{\rm SM}  +   {\cal O}(h_{\mu \nu}^2) \right], 
\label{int}
}
where
\dis{
T^{\mu \nu}_{\rm SM} = \left.-2 \frac{\partial {\cal L}_{\rm SM}}{\partial g_{\mu \nu}} + g^{\mu \nu} {\cal L}_{\rm SM} \right|_{g_{\mu \nu} = \eta_{\mu \nu}}.
}

After integrating out the extra dimension from Eq. (\ref{Sh}) and using Eq. (\ref{int}) for interaction terms, the 4D effective action for the massless graviton and the KK gravitons in the Minkowski background comes out as
\dis{
{\cal S} = \int d^4 x \left[ -\frac12 (\partial_\rho h_{\mu \nu}^{(0)})^2  -\frac{h_{\mu \nu}^{(0)}}{M_P} T_{\rm SM}^{\mu \nu} -\sum_{n=1}^\infty \left(\frac12 (\partial_\rho h_{\mu \nu}^{(n)})^2 + \frac12 m_n^2 (h_{\mu \nu}^{(n)})^2+  C_n \frac{h_{\mu \nu}^{(n)}}{M_5} T_{\rm SM}^{\mu \nu} \right) \right]\,,
\label{lag}
}  
where $m_n$ is given in Eq. (\ref{KKm}), and  
\dis{
M_P = \sqrt{\frac{M_5^3}{k} (e^{2k\pi R} -1) }, \label{MpM5}
}
\dis{
C_n 
\begin{cases}
\simeq (-1)^{n-1} \sqrt{\frac{p}{M_5}} ,&  p>0 \\ 
=\frac{1}{\sqrt{M_5 \pi R}}\frac{n}{\sqrt{k^2 R^2 + n^2}}, & p=0\\
 \simeq \sqrt{\frac{\pi |p|}{M_5}} \frac{e^{-k \pi R}}{\Gamma(k/|p|)} \left[\frac{\pi}{2} \left(n-\frac14 + \left|\frac{k}{2p}\right|\right)  \right]^{\frac{k}{|p|}-\frac12} ,&  p<0  \,.
 \end{cases} \label{KKC}
}

Eq. (\ref{MpM5}) shows that the 5D Planck scale $M_5$ can be exponentially smaller than the 4D Planck scale $M_P$, which provides a solution to the electroweak hierarchy problem if $M_5$ is around the electroweak scale. This exponential hierarchy $\sim e^{(k_1+k_2/2) \pi R}$ between $M_5$ and $M_P$ stems from two factors: a large extra dimension with size $L \sim e^{k_2 \pi R}/k_2$ and the power-law warping $(k_2 L)^{k_1/k_2} \sim e^{k_1 \pi R}$ as can be seen from the metric (\ref{metric1'}). Therefore, GLD with $k_2 \neq 0$ including LD addresses the hierarchy problem by both a large extra dimension (as ADD) and a warp factor (as RS). 

The masses and couplings of the KK gravitons in GLD with $p<0$ show a particularly different parametric behavior compared to the conventional models such as ADD, RS, and LD. 
Taking $k\sim p \sim M_5$ for simplicity, Eq. (\ref{KKm}) and Eq. (\ref{KKC}) shows that $m_n \sim n M_5$ and $C_n \lesssim {\cal O}(1)$ for $p \geq 0$ as in RS and LD. On the other hand, for $p<0$, the KK graviton masses are  $m_n \sim n M_5 (M_5/M_P)^{|p|/k}$ as in ADD with $2k/|p|$ flat extra dimensions, but the KK graviton couplings are $C_n \sim (M_5/M_P) \times n^{(k/|p| -1/2)}$ which is clearly different from ADD where $C_n = M_5/M_P$. This difference can be significant for heavy KK states with large $n$. As we will show in section \ref{sec:EAlimits}, this feature allows better sensitivity on $M_5$ at future colliders such as CLIC compared to ADD models due to larger cross section for heavy KK states.

\section{Signatures of extra dimensions}\label{sec:searches}
In this section, we present the observable characteristics of RS, LD, and GLD models, which will be further investigated in present and future experiments, as well as in astrophysical observations. 
Since the dominant decay channels of KK gravitons typically involve SM states, with exception of the $\lambda \gtrsim 1$ GLD models, as we discuss further, we examine their visible decays in two distinct regimes: short-lived, with decay lengths below $1\, m$, and long-lived, with decay lengths exceeding $1\, m$. The former regime is relevant to collider searches, while the latter is pertinent to beam dump experiments, astrophysical, and cosmological observations. Since these two regimes are characterized by different physical features, combining the two allows for an effective coverage of a large part of the parameter space. 

\subsection{Short-lived regime}\label{sec:short_LLP}
In this regime KK gravitons decay with decay lengths of $d\lesssim 1\,m$. They can be searched at the LHC \cite{Antoniadis:2001sw,Giudice:2004mg,Giudice:2017fmj} or at the upcoming $e^+e^-$ colliders like Future Circular Collider (FCC-ee) \cite{FCC:2018evy} and Compact Linear Collider (CLIC) \cite{CLIC:2016zwp}. Since the branching ratios of decays into a pair of photons or leptons may be sizable, and constitute a clean experimental channel, we will mainly focus on them. Moreover, a lepton version of the FCC, is planned to be constructed before hadron collider.
On the other hand, in section \ref{sec:EAlimits}, we will also present results for KK gravitons decaying into any visible SM states, mainly into heavy quarks and gluons, which constitute the majority of the decay widths.

\paragraph{The LHC}
At the LHC, the $n$-th KK graviton can be produced by gluon fusion or through interactions with quarks. The total cross section is given by \cite{Giudice:2017fmj} 
\beq
\sigma_n = \frac{\pi}{48 \Lambda_n^2} \left(3\mathcal{L}_{gg}(\hat{m}_n^2) + 4 \sum_q \mathcal{L}_{qq}(\hat{m}_n^2) \right)\,,
\label{eq:sigma_LHC}
\eeq
where $1/\Lambda_n = C_n/M_5$ is the KK graviton coupling to the SM fields as given in Eq. (\ref{KKC}), $q$ are quark fields, $\mathcal{L}(\hat{s})$ are parton luminosities,
\beq
\mathcal{L}(\hat{s}) = \frac{\hat{s}}{s} \int_{\frac{\hat{s}}{s}}^{1} \frac{dx}{x} f(x) f\left(\frac{\hat{s}}{x s}\right)\,,
\label{eq:pdf_LHC}
\eeq
where $s=13\,\tev$, and $f(x)$ are the parton distribution functions, which we take from LHAPDF6 data~\cite{Martin:2009iq,Clark:2016jgm,Buckley:2014ana}.

The number of di-photon events is:
\beq
N = \mathcal{L} \times \sum_n \sigma_n \times \mathrm{BR}(G_n \to \gamma\gamma),
\label{eq:sigma_LHC}
\eeq
where $G_n$ denotes the $n$-th KK graviton mode, and we sum over the modes decaying inside the ATLAS/CMS detectors.
In order to determine the exclusion limits from the LHC, we follow the discussions in \cite{Baryakhtar:2012wj,Giudice:2017fmj} and use the data from \cite{ATLAS:2017fih,CMS:2017yta}.

\paragraph{Electron-positron colliders}
At the lepton colliders, we calculate the production cross section of KK gravitons in association with a photon or $Z$ boson, $e^+e^- \to G\gamma$ and $e^+e^- \to GZ$, respectively.\footnote{We also took into account the resonant production $e^+e^- \to G$, which we found to be subdominant, see Eq. (\ref{eq:amp2_eeG}).}
For the former process, we use results from \cite{Giudice:1998ck,Han:1998sg}, while for the latter process, we calculated the averaged amplitude squared, given by Eq. (\ref{eq:amp2_eeGZ}).

The number of decays is given by luminosity multiplied by the total production cross section and the branching ratios into the relevant SM states (mainly a pair of photons or leptons):
\beq
N = \mathcal{L} \times \sum_n \sigma\left(e^{+} e^{-} \rightarrow X G_n\right) \times \mathrm{BR}_n.
\label{eq:NoE_short}
\eeq
The effective production cross section is given by
\beq
\sigma\left(e^{+} e^{-} \rightarrow X G_n\right)=\int d \Omega \, \frac{d \sigma\left(e^{+} e^{-} \rightarrow G X\right)}{d \Omega}\, \left(1 - e^{-R/L_G^{\perp}(\theta)}\right)\,,
\label{eq:master_short}
\eeq
where $L^{\perp}_G=c \tau \gamma \beta \sin\theta$ with $\beta \equiv v/c$ and $\gamma\equiv 1/\sqrt{1-\beta^2}=(s-m_{X}+m_G^2)/(2m_G \sqrt{s})$ for $X=\gamma$, $Z$, and $\theta$ is the angle measured from the collider axis.
With regard to the FCC-ee and CLIC, we assume the following data in our analysis: $R_{\mathrm{FCC}}=0.5$ m , $\mathcal{L}_{\mathrm{FCC}_1}=145\,$ab$^{-1}$ for $\sqrt{s}= 91\,\gev$, $\mathcal{L}_{\mathrm{FCC}_2}=20\,$ab$^{-1}$ for $\sqrt{s}= 161\,\gev$, $\mathcal{L}_{\mathrm{FCC}_3}=\,5$ab$^{-1}$ for $\sqrt{s}= 250\,\gev$ \cite{FCC:2018evy}, and $R_{\mathrm{CLIC}}=0.6$ m, $\mathcal{L}_{\mathrm{CLIC}}=3\,\mathrm{ab}^{-1}$ for $\sqrt{s}= 3000\,\gev$ \cite{CLIC:2016zwp}.

To determine prospects of searches at lepton colliders, we largely followed the approach presented in Ref.~\cite{Bauer:2018uxu}, which assumed zero background for the $e^+e^- \to \mathrm{ALP}+\gamma \to 3\gamma$ search, which is consistent with the results of \cite{Knapen:2016moh} that determined the LEP \cite{OPAL:2002vhf} limits for photon-coupled ALP.
While a sizable SM background from light hadrons and $Z$ boson decays into two photons is expected \cite{dEnterria:2023wjq}, such events will be associated with the known mass and width of the decaying SM particle. 
Therefore, they can be vetoed, and their masses will be excluded from the projections for any BSM species also decaying into two photons.
In fact, the recent reanalysis of the FCC sensitivity to the photon-coupled ALP \cite{RebelloTeles:2023uig} validated the results of \cite{Bauer:2018uxu} obtained under zero-background assumption, which also justifies our approach.

\subsection{Long-lived regime}\label{sec:long_LLP}
When $d\gtrsim 1\,m$, the long-lived particles (LLPs) are effectively collider-stable, escaping the detectors before decaying, leaving a missing-energy signature. 
Following \cite{Ovchynnikov:2023wgg}, we checked that the invisible signature $e^+e^- \to \gamma + \mathrm{inv.}$ at FCC-ee does not impose leading constraints on any GLD model.

In fact, the KK gravitons decay predominantly into visible SM states, except the $\lambda\gg 1$ models in the region of the parameter space with light, $m_G\lesssim 10\,\mev$, KK gravitons that decay mostly into other gravitons. 
In that case, however, the terrestrial experiments are typically not competitive with astrophysical bounds, as discussed in section \ref{sec:EAlimits}.
Therefore, beam dump experiments are more suitable to probe the long-lived regime of $\lambda\lesssim 1$ GLD, in particular the RS and LD backgrounds, since there the KK gravitons decay predominantly into visible SM states.

\paragraph{Beam dump experiments} use a high-energy, $\sim 100\,\gev$, electron/proton beam hitting a dense target, producing LLPs in the forward direction. The decay vessel, where LLPs decay into a pair of photons or charged SM states, is situated at a distance $\sim 100\,m$ from the production point, allowing to study LLP signature with highly displaced vertex. 
We investigate visible decays of KK gravitons at beam dump experiments or forward LHC detectors: DUNE \cite{DUNE:2020ypp}, FASER2 \cite{FASER:2018eoc}, MATHUSLA \cite{Curtin:2018mvb}, NA62~\cite{Dobrich:2018ezn}, SeaQuest~\cite{Berlin:2018pwi}, and SHiP \cite{Alekhin:2015byh}; for each of them, we model their characteristics as given in Tab.~1 of \cite{Jodlowski:2023yne}.

The number of LLP decays occurring within the detector positioned between $[L_{\mathrm{min}},L_{\mathrm{max}}]$ from the production point is
\beq
N = \sum_{E,\theta} N_{\mathrm{LLP}}(E,\theta) \times \left(e^{-L_{\mathrm{min}}/d(E)}-e^{-L_{\mathrm{max}}/d(E)}\right),
\label{eq:master_long}
\eeq 
where $N_{\mathrm{LLP}}(E,\theta)$ is the LLP spectrum, and the second factor corresponds to the probability of decays taking place inside a detector, and $d(E)=c\tau \gamma \beta$ is the decay length.
Given that beam dump experiments are able to perform the search for high-energy displaced decays with essentially no background, guaranteed by large separation between LLP production and decay point, use of magnetic field and hadronic absorbers, we plot exclusion bounds based on $N=3$ events.

The LLP spectrum, which describes the energy and angular distribution, is determined by rare meson decays, gauge boson (photon) fusion or conversion due to scattering or interaction with electromagnetic field, and bremsstrahlung, while other contributions are subdominant. 
The main production mode of $\lesssim \gev$ KK graviton $G_k$ at beam dumps is Primakoff conversion, whose cross section is $\sigma_{\gamma N \to G_k N} \sim \alpha_{\mathrm{EM}} Z^2/(2 \Lambda_k^2)$, where $Z$ is the nucleus atomic number, and $\Lambda_k$ is the coupling to two photons \cite{Jodlowski:2023yne}. 
Moreover, KK gravitons can be produced via $B \to K+G_k$ decays.
This mode of $G_k$ production was neglected in \cite{Bijnens:2000wj,Giudice:2004mg} because the energy momentum tensor involving $(B,K)$ states is diagonal at leading order, and as a result this decay channel is loop-suppressed. 
In fact, explicit computation of such transition was done in \cite{Degrassi:2008mw}, and we checked that it is indeed subdominant with respect to the Primakoff conversion.

\paragraph{FCC LLP mode}
In $e^+ e^-$ collisions at high energies, in particular in the LLP mode of FCC-ee \cite{Blondel:2022qqo,Drewes:2022rsk}, it is also possible to produce KK gravitons $G_n$ from the decays of on-shell $Z$ bosons, $Z \to G f\bar{f}$, where $f\bar{f}$ denotes SM fermions (mostly $b\bar{b}$ quarks).
After KK graviton production, they decay into two photons.
This is not a background-free search and to obtain the FCC-LLP bound, we follow the discussion in \cite{Drewes:2022rsk,Ovchynnikov:2023wgg,Blondel:2022qqo}. 
We note that since the gauge bosons are localized on the same brane, the $Z-\gamma-G_n$ coupling vanishes at tree level. Moreover, this coupling induced at 1-loop \cite{Nieves:2005ti,Allanach:2007ea} is suppressed by $\alpha_{EM}^2$, and does not lead to competitive bounds.
As a result, the production of KK gravitons in this mode is phase-space suppressed, $N_{G_n}=N_Z \times \mathrm{BR}(Z\to G_n f\bar{f})$, where $N_Z=2.5 \times 10^{12}$, while the averaged amplitude squared for this process is given by Eq. (\ref{eq:amp2_ZGee}).

\paragraph{Astrophysical bounds}
In addition to the terrestrial searches, the long-lived regime can be probed by Big-Bang nucleosynthesis (BBN) \cite{Kawasaki:2004qu}, neutron stars (NS) observations (gamma rays and excess heating) \cite{Hannestad:2003yd,Casse:2003pj}, and SN1987 \cite{Ishizuka:1989ts,Dev:2020eam,Hannestad:2003yd}. 
We follow discussion in \cite{Hannestad:2003yd,Casse:2003pj,Baryakhtar:2012wj}, except for the NS bounds, where we believe these limits determined in \cite{Hannestad:2003yd,Casse:2003pj} are in fact weaker by a factor $\sim 2$ - see appendix \ref{app:astro_bounds_ADD}. This affects only the $\lambda \gtrsim 1$ GLD models, since only very long-lived KK gravitons can survive for timescales relevant to NS.

In particular, we note that the limit from the NS excess heating, $M_5>1.6\times 10^5\,\tev$, \cite{Hannestad:2003yd} does not correspond to the stated limit on the excess luminosity $L< L_{\mathrm{max}}= 10^{-5}\, L_{\odot}$, but instead corresponds to $L_{\mathrm{max}}= 4\times 10^{-8}\, L_{\odot}$.
In our analysis, we assume that this discrepancy is just a typo in the equation for $L_{\mathrm{max}}$, and that the correct limit on $M_5$ corresponds to $L_{\mathrm{max}}=10^{-5}\, L_{\odot}$, which is also consistent with~\cite{Larson:1998it}. 
This means that the limit on $M_5$ for the ADD scenario with $n=1$ extra dimensions is weaker by factor $\sim 10$,  see the fourth row and the second column in Tab. \ref{tab:tab1}.

\section{Experimental and astrophysical limits  }\label{sec:EAlimits}
The GLD with $\lambda^2 >0$ has an approximate dilatonic shift symmetry,\footnote{For $\lambda^2=0$, the Randall-Sundrum background, the low $k/M_5$ ratio is not natural, since the dilaton is decoupled. However, RS with small curvature $k$ can be considered as a deformation of the ADD that significantly opens up the allowed parameter space by modifying the KK spectrum, which also leads to interesting collider signatures \cite{Giudice:2004mg,Kisselev:2005yn}.} which is softly broken by terms dependent on the curvature $k$. Therefore, the regime of $k$ small compared to $M_5$ is technically natural. Since it involves light, weakly coupled KK states, it is quite difficult to probe and gathered relatively little phenomenological attention, although see \cite{Giudice:2004mg,Cox:2012ee,Baryakhtar:2012wj}. 
Since the phenomenology of RS and LD at the LHC has been extensively studied \cite{ATLAS:2017fih,CMS:2017yta,ATLAS:2023hbp,Giudice:2017fmj}, we focus on the small curvature regime for these backgrounds. Adapting the discussion of \cite{Giudice:2017fmj} to the RS, see also section \ref{sec:short_LLP}, we update the LHC bounds on this scenario as well. On the other hand, since phenomenology of GLD with $\lambda^2 >1$ has not been studied before, we provide its comprehensive analysis according to the discussion in section \ref{sec:searches}, taking benchmark points $\lambda^2=2, 3, 4, 6, 10$, and $100$.

\begin{figure}[th]
    \includegraphics[scale=0.36]{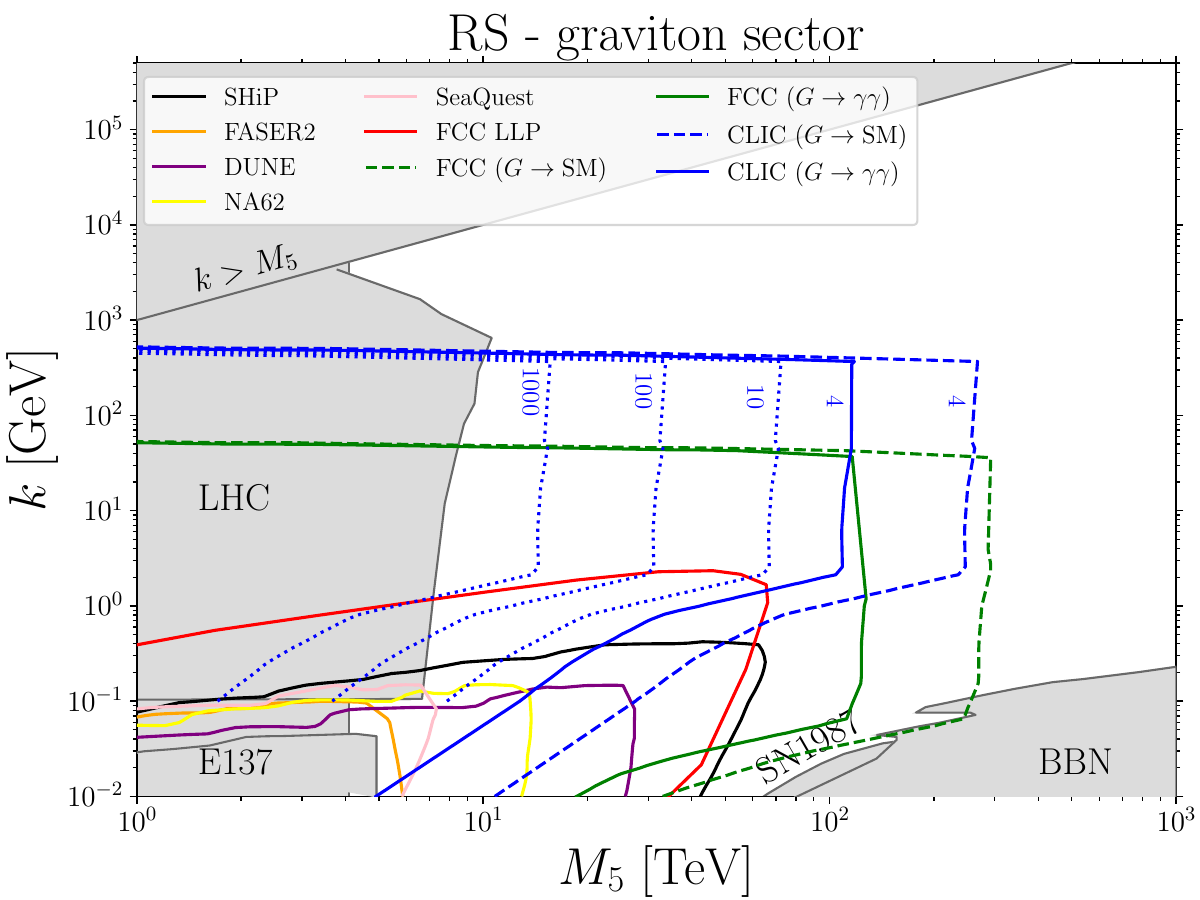}
    \hspace{0.1cm}
    \includegraphics[scale=0.36]{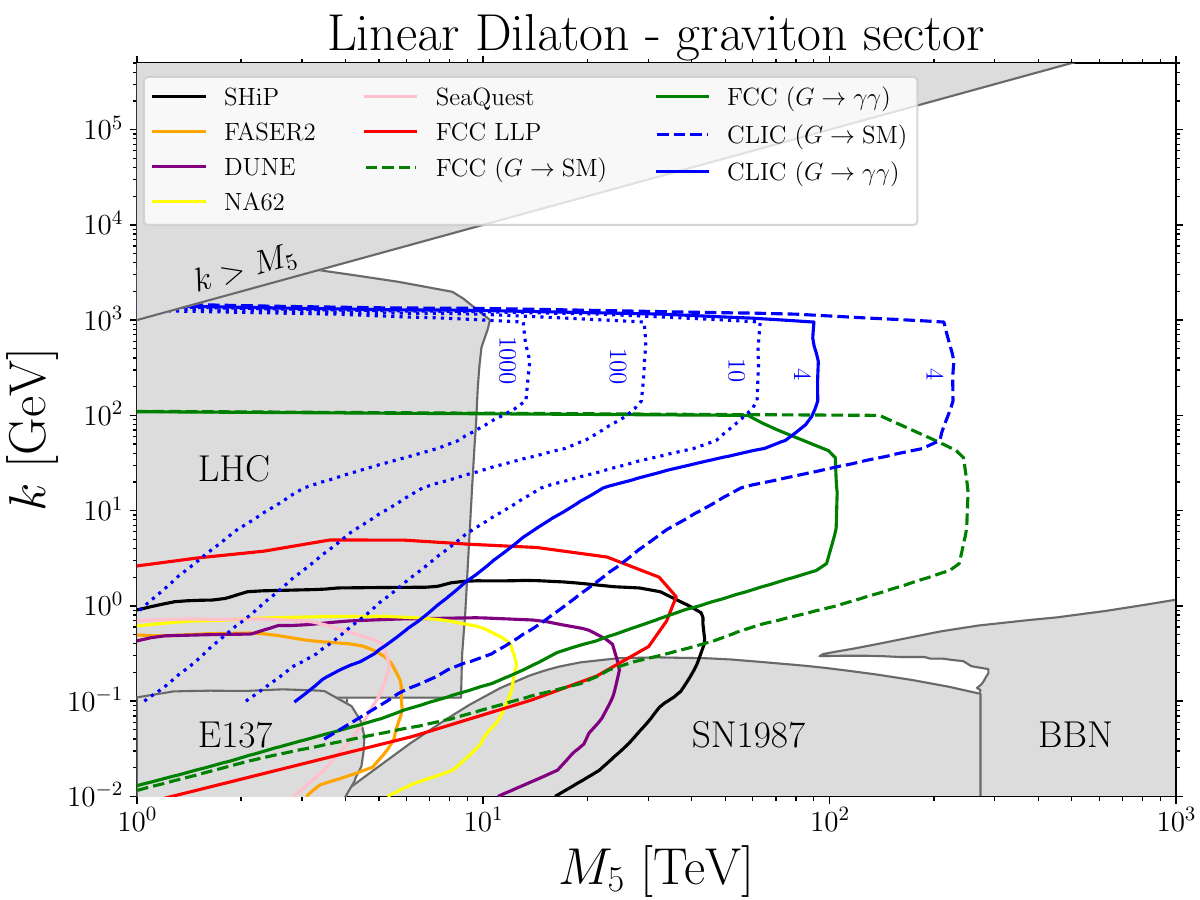}\\
    \\
    \includegraphics[scale=0.36]{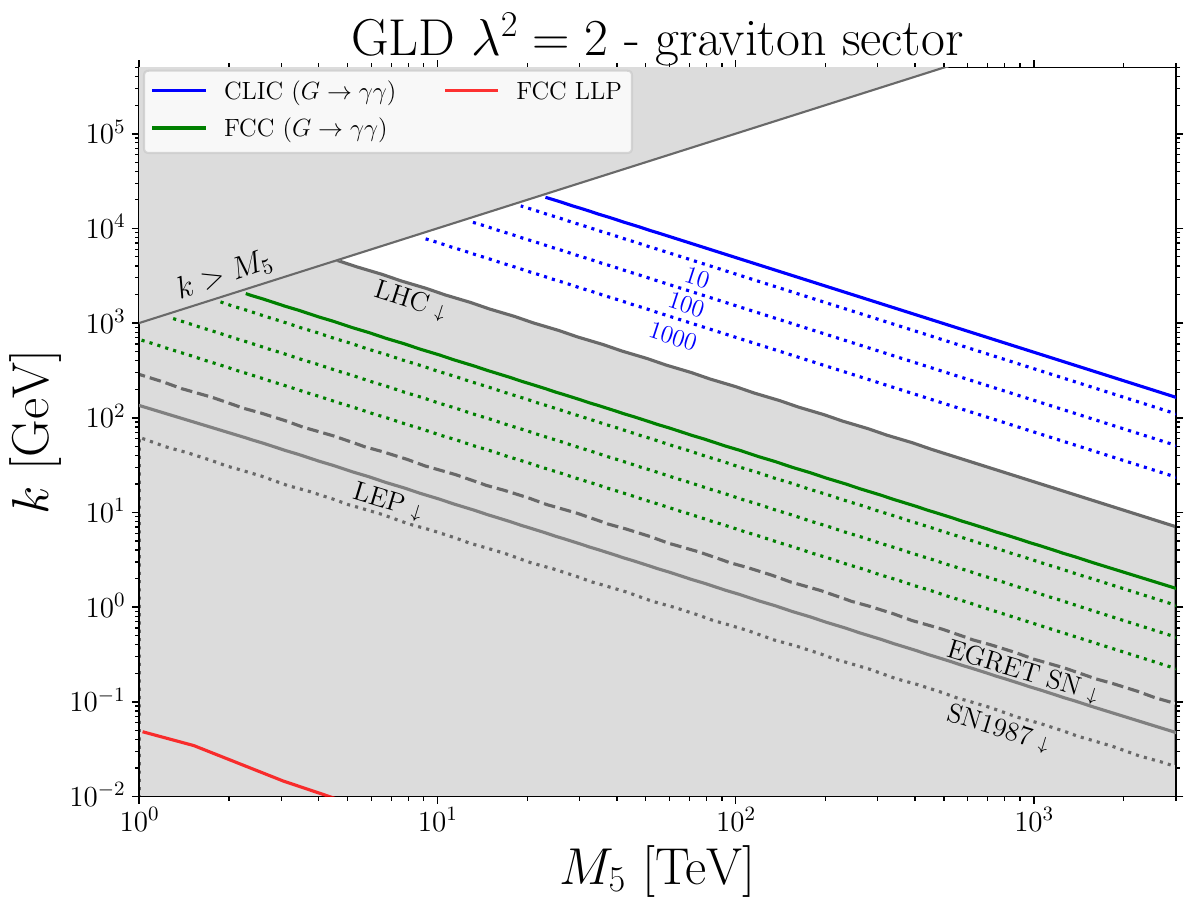}
      \hspace{0.1cm}
    \includegraphics[scale=0.36]{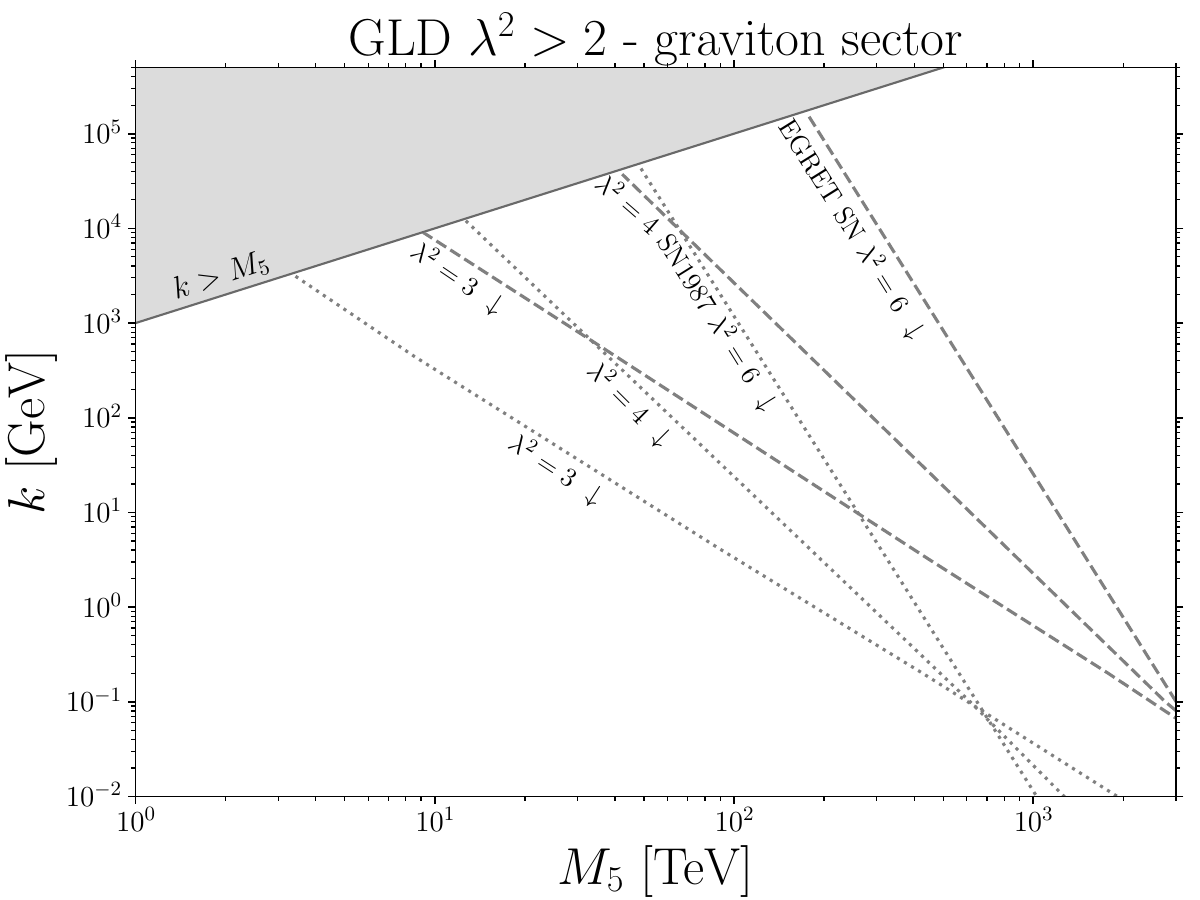}
    \caption{
            Limits and projections for the extra dimensional models obtained within the GLD. The results shown correspond to: the Randall-Sundrum model (\textit{top left}), linear dilaton background (\textit{top right}), generalized linear dilaton background with $\lambda^2 >1$ (\textit{bottom}). The RS and LD background correspond to $\lambda^2=0,1$, respectively. The solid colorful lines indicate projections for future experiments, while grayed out regions are already excluded. The blue numbers $4,10,100,1000$ denote the number of events at CLIC. 
            The final sensitivity projection for CLIC and FCC corresponds to blue and green solid lines, respectively. 
            The \textit{bottom left} plot contains collider and astrophysical exclusion bounds - SN1987 (dotted) and EGRET SN (dashed) - for $\lambda^2=2$, according to the legend. We also show astrophysical bounds on larger values of $\lambda^2=3, 4, 6$ (\textit{bottom right}). 
        }
\label{fig:Gen_Clock}
\end{figure}

In fig. \ref{fig:Gen_Clock}, we show the current bounds and projected sensitivity of upcoming experiments to the previously introduced benchmarks of the GLD. 
The top panels correspond to RS and LD, $\lambda^2=0,1$, respectively, while results for GLD with $\lambda^2>1$ are shown in the bottom panels. 
In all panels the gray-shaded regions indicate the currently excluded parameter space or theoretically disallowed parameter space with $k>M_5$, while the projected sensitivities of future experiments and astrophysical constraints are represented by solid, dashed, or dotted lines.

For RS ($\lambda^2=0$) and LD ($\lambda^2=1$), the KK gravitons couplings allow for their sizable production at the LHC and lepton colliders, only weakly dependent on the masses of the KK gravitons.
For FCC-ee and CLIC we present two projections: assuming all visible decay products of $G$ - mostly gluons and quarks - can be detected (dashed line), and only considering decays to a pair of photons (solid line). 
For the latter types of decays, we also show lines corresponding to 10, 100, and 1000 events, in case the zero-background assumption is not satisfied.\footnote{See the discussion in section \ref{sec:short_LLP}, especially the last paragraph.}
For the RS and LD backgrounds, the $k\gtrsim 1\,\gev$ region, which is currently covered up to $M_5\sim 10\,\tev$ by the LHC \cite{ATLAS:2017fih,CMS:2017yta,ATLAS:2023hbp,Giudice:2017fmj}, will be extended up to $M_5 \sim 200\,(100)\,\tev$ by CLIC and FCC-ee searches for $G$ decays into a pair of any visible SM states (photons or leptons). 
The presented limits correspond to combined searches using $e^+e^- \to G\gamma$, $e^+e^- \to GZ$, and resonant production for both experiments. 
Because of the larger center of mass energy of CLIC and larger detector, it covers larger values of $k$ than FCC-ee, which in turn covers most of the short-lived regime.
The long-lived regime is covered by the BBN, SN1987, FCC-ee LLP, and beam dump experiments or forward LHC detectors: DUNE, FASER2, MATHUSLA, NA62, SeaQuest, and SHiP. The strongest terrestrial present constraints come from NuCal \cite{Blumlein:1990ay} and E137 \cite{Bjorken:1988as}. We obtained a similar but slightly weaker exclusion line derived for E137 than the limits found in \cite{Baryakhtar:2012wj}, which we believe is due to the neglect of the invisible decays (overwhelmingly into a pair of neutrinos, which constitute roughly $40\%$ of the total decay width when $m_G\lesssim 2m_e$) of KK gravitons in \cite{Baryakhtar:2012wj}. 

In contrast to RS and LD, the GLD models with $\lambda^2>1$ are characterized by continuum of densely packed KK states and small KK-graviton couplings similarly to the ADD model, while the KK-graviton couplings are proportional to some power of the KK-graviton masses. This gives rise to a interestingly distinctive feature that these models predict both a sizable flux of collider stable, light ($m\lesssim 30\,\mev$) KK gravitons, which lead to strong astrophysical bounds coming from SN observations, and comparably strong flux of heavy, $m\gtrsim 1\,\tev$, KK gravitons decaying into visible SM states, which can be probed by colliders.
As shown in the bottom left panel of Fig.~\ref{fig:Gen_Clock}, the LHC has excluded the parameter space up to $M_5 =k \sim 5\,\tev$, in addition to the tail occurring from the scaling of the total production cross-section of the KK gravitons  $\sigma_{\rm tot} \sim \sum_n C_n^2/M_5^2 \sim k (\sqrt{s}/k)^q / (q M_5^3)$ with $q=(\lambda^2+2)/(\lambda^2-1)$ for the center of mass energy of the collider $\sqrt{s}$. 
Moreover, CLIC will improve upon the LHC bounds by probing the parameter space up to $M_5 =k \sim 20\,\tev$, while FCC, due to its smaller center of mass energy, cannot produce heavy on-shell KK gravitons, resulting in the improvement by about an order of magnitude  upon the limit set by LEP.
On the other hand, displaced vertex searches for $\sim$GeV KK gravitons at SHiP and FCC LLP mode turn out to set weaker limits than the astrophysical bounds derived from SN1987 and other SN, which are sensitive to KK gravitons with masses $\lesssim 30\,\mev$.
Finally, for $\lambda^2 \gtrsim 3$, the couplings of KK gravitons are too feeble to allow setting collider bound that would be competitive with the astrophysical limits.
Therefore, in the bottom right panel of Fig. \ref{fig:Gen_Clock}, we  only show the relevant astrophysical bounds - the SN limits for $\lambda^2=3, 4$, and $6$. 
The value $\lambda^2=6$ corresponds to the smallest parameter that can be realized in heterotic M-theory \cite{Im:2018dum}, and the resulting SN limits (denoted as SN1987 and EGRET SN) approach to the $N=1$ ADD limits as $k\rightarrow 0$, $M_5\gtrsim 740 \,\tev$ and $M_5\gtrsim 3400 \,\tev$, respectively - see Table \ref{tab:tab1}.

For even larger values of $\lambda^2$, the decays of KK gravitons into other KK states, lighter KK gravitons or KK dilatons, are important. To compute the corresponding decay widths, we use the results of Ref.~\cite{deGiorgi:2021xvm}, in particular Eq. A.6 therein, which was confirmed in, \eg, Ref.~\cite{Chivukula:2024nzt}.
For the LD background, Ref.~\cite{Giudice:2017fmj} found that the $G_n \to \sum_{k,l}G_k G_l$ decays dominate over the decays into SM states in the $k\ll M_5$ regime - see Fig. 26 and Appendix D therein.
On the other hand, in Appendix A of Ref.~\cite{deGiorgi:2021xvm} it is stated that Eq. A.6 therein, which seems to be the correct expression since it agrees with the results (amplitude unitarization) of Ref.~\cite{Bonifacio:2019ioc}, does not match the result of Ref.~\cite{Giudice:2017fmj}.
In light of that, in our analysis, we follow Ref.~\cite{deGiorgi:2021xvm} to compute the inner tower decays of KK gravitons. 

\begin{figure}[h]
    \centering
    \includegraphics[scale=0.4]{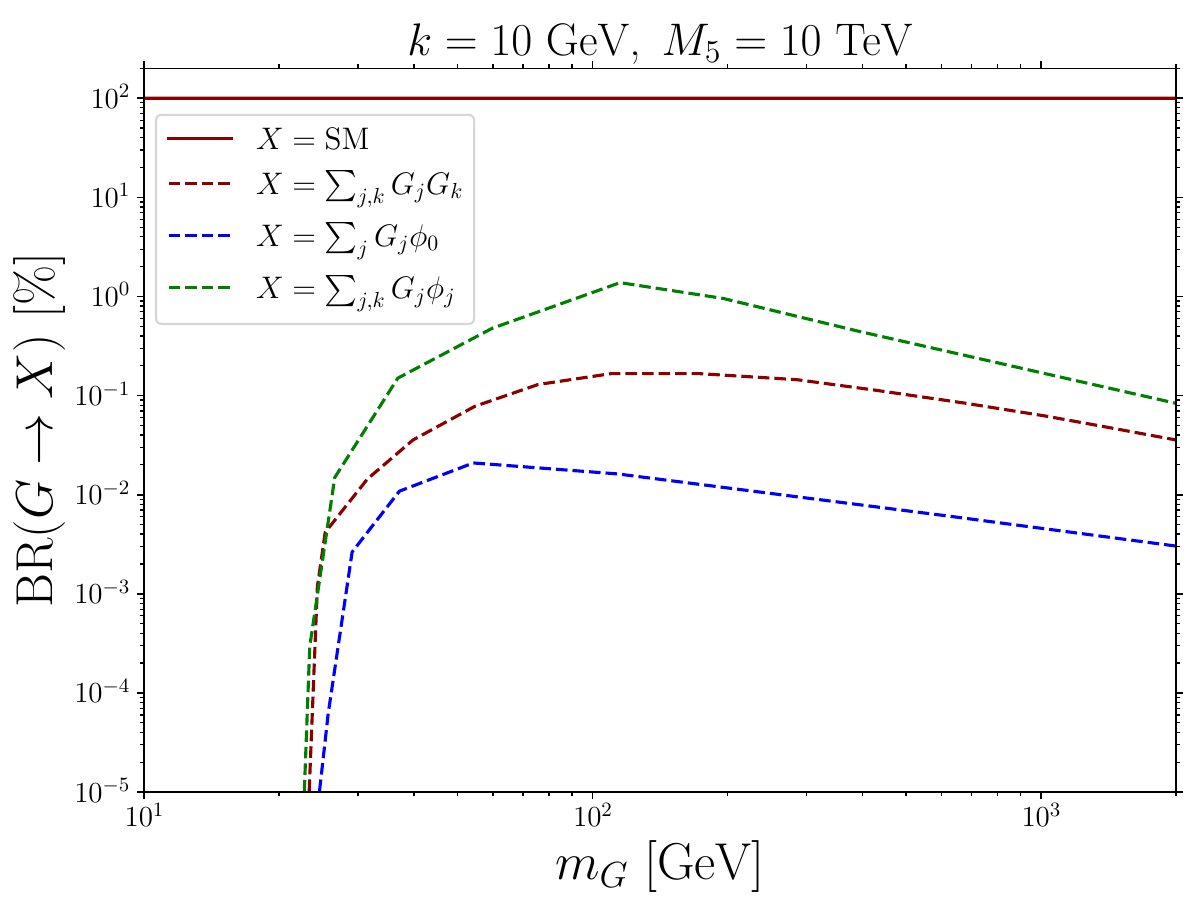}
    \caption{
        Branching ratios of KK gravitons as a function of their mass for LD background. Note that the decays into SM particles (red solid line) dominate over decays within the KK tower - into a pair of lighter gravitons (red dashed line) or into a single KK graviton and radion (blue dashed line) or any KK-dilaton (green dashed line). We fixed $k$ and $M_5$ as shown in the plot title for comparison with Fig. 26 of \cite{Giudice:2017fmj} - see also the remarks in Appendix 1 in \cite{deGiorgi:2021xvm}. 
        }
\label{fig:BR_Gn}
\end{figure}

\begin{figure}[h]
    \includegraphics[scale=0.36]{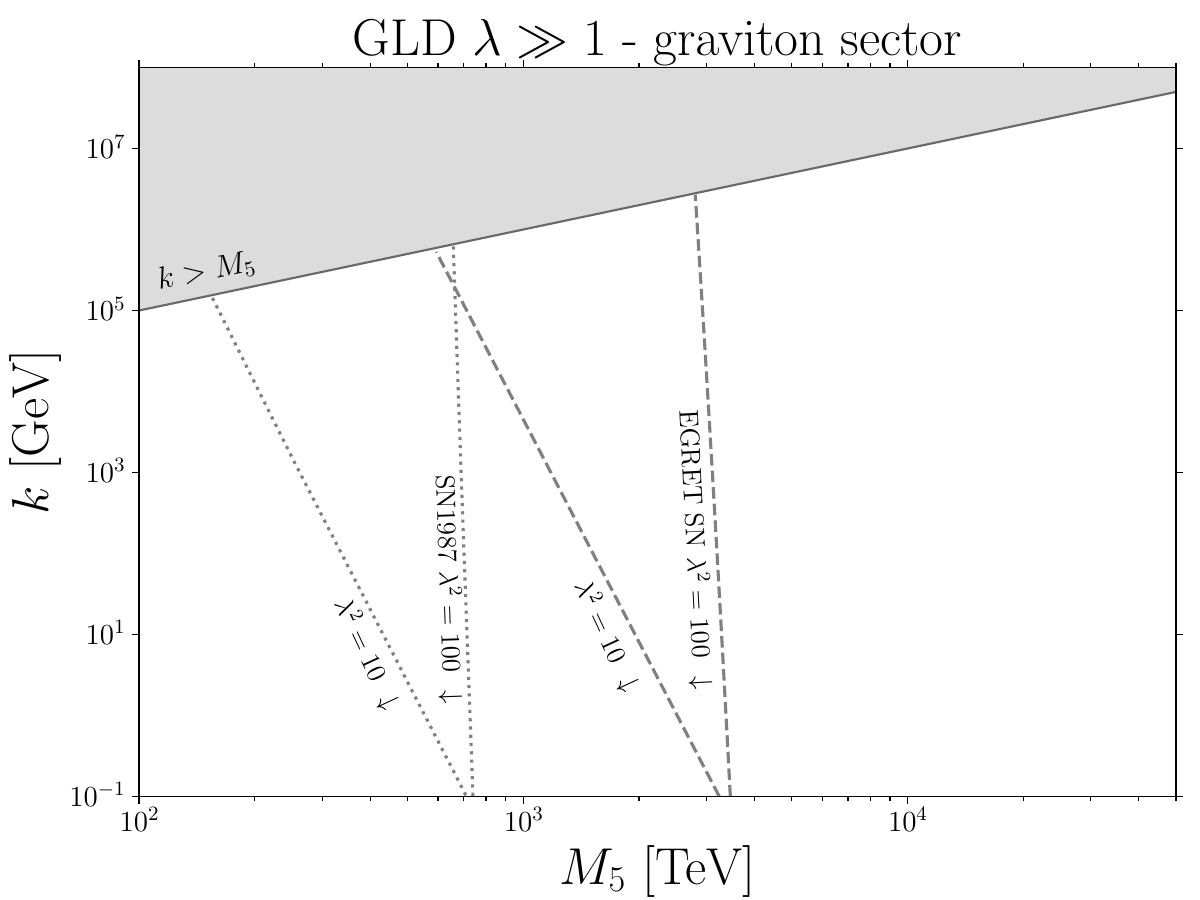}
    \hspace{0.1cm}
    \includegraphics[scale=0.36]{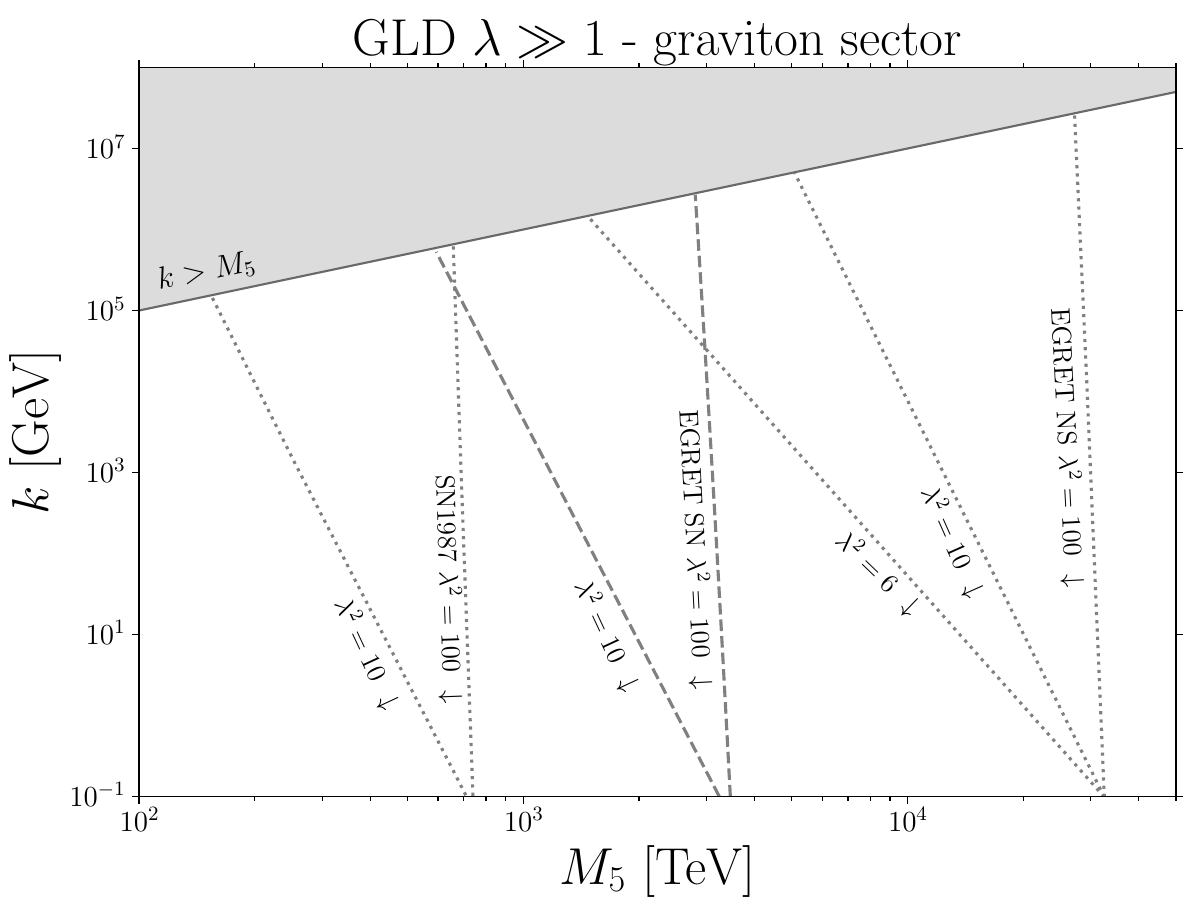}
    \caption{
        Limits on GLD with $\lambda^2\gg 1$ with (left) and without (right) taking into account the KK-tower decays of KK gravitons, which significantly shorten their lifetimes. As a result, light KK gravitons, which are produced in supernovae, decay before the neutron stars are formed and the corresponding bounds are evaded, as seen from the comparison of the left and right panels. On the other hand, the SN bounds are not affected, since the KK gravitons are effectively stable on this timescale. 
        }
\label{fig:Gen_Clock_compare}
\end{figure}

In Fig. \ref{fig:BR_Gn}, we present the results for the LD background, where we fixed the curvature $k$ and $M_5$ to the same values as in Fig. 26 of \cite{Giudice:2017fmj}. 
Our result is indeed different from \cite{Giudice:2017fmj} - the KK graviton decays within the KK tower are in fact negligible for any signature - both terrestrial experiments and astrophysical observations.
On the other hand, for GLD with $\lambda^2>1$, these decays are important, since astrophysical bounds strongly depend on the lifetimes of light KK gravitons.
In fact, the strongest astrophysical bounds come from NS observations, which assume that LLPs have lifetimes at least as large as the corresponding neutron star.
For flat extra dimensions, the KK-tower decays are forbidden and the KK gravitons have only feeble interactions with the SM, thus they are extremely long-lived.
This regime is recovered for $\lambda^2\gg1$ when $k\ll M_5$, where we recover the following ADD limits~\cite{Hannestad:2003yd,Casse:2003pj}, see also appendix \ref{app:astro_bounds_ADD}: $M_5>740\,\tev$, $M_5>3400\,\tev$, and $M_5>1.2\times 10^4\,\tev$, obtained by SN1987, EGRET SN, and NS excess heating, respectively.
On the other hand, in GLD with non-zero curvature, KK gravitons can have quite sizable interactions with other KK states, leading to much suppressed lifetimes, which relax the NS bounds.
We show our results in Fig. \ref{fig:Gen_Clock_compare} for $\lambda^2=6, 10, 100$, where one can see that taking into account the invisible decays (the left panel) leads to weaker bounds.
Unfortunately, the SN1987 and EGRET SN limits - which  are the same on both panels - are sufficiently strong to exclude GLD with $\lambda^2\gg1$ as a natural solution to the hierarchy problem.

\section{Conclusions\label{sec:conclusions}}
In this work, we have examined the phenomenology of power-law warped extra dimensions realized by General Linear Dilaton models as well as updating the phenomenological prospects of the two conventional models, Randall-Sundrum and Linear Dilaton model.
We updated or determined the current bounds on these models using the LHC searches, beam dumps, and astrophysical observations. Furthermore, we determined the sensitivity of future lepton colliders FCC-ee and CLIC, which will probe the $c \tau \lesssim 1\,m$ part of the parameter space for these models. As a result, the projected sensitivities extend the coverage up to $M_5 \sim 200\, \tev$, improving the current bounds by a factor $\sim 20$.

We also studied the long-lived regime, $c \tau \gg 1\,m$, particularly the region where the curvature scale $k$ is small compared to the 5D Planck mass $M_5$. The low curvature regime is motivated due to an approximate shift symmetry of the dilaton. This regime can also be interpreted as deformations of the flat ADD scenario with only one additional space dimension, which avoids stringent bounds due to warping.

We determined the prospects of the currently running or approved beam dump experiments like FASER, NA62, and SHiP, and the proposed DUNE, MATHUSLA, and SeaQuest, for detecting decays of $\sim$sub-GeV KK gravitons.
We also demonstrated their synergy with the searches at the LLP mode of FCC-ee utilizing the $Z$ boson decays. 
Consequently, future experiments will therefore comprehensively test both the short and long-lived parameter space of the GLD models.

Generalizing the observations made in \cite{Giudice:2017fmj} for the LD, GLD with $\lambda^2 >1$ leads to a different phenomenology from both ADD and RS models due to its different structure of KK states - their masses and couplings. 
In particular, KK gravitons are characterized by a compressed spectrum leading to cascade decays into high-energy SM particles. 
If such a signal emerges in searches for long-lived particles, it could be a smoking gun signature of a power-law warped extra dimension. 
Moreover, we studied the corresponding astrophysical bounds, where we found that the observations of SN1987 and neutron stars can place the most stringent bounds for GLD with $\lambda^2 \gtrsim 1$. In particular, our revised astrophysical bounds on ADD scenario with $1$ extra dimension (corresponding to $\lambda^2 \to \infty$), are weaker than stated in the literature~\cite{Hannestad:2003yd,Casse:2003pj} by a factor $\sim 2$.
Our results may therefore motivate further phenomenological and experimental efforts to investigate the GLD.

\section*{Acknowledgments}

KJ thanks Stephen Angus, Brian Batell, Sungwoo Hong, Fotis Koutroulis, Miguel Montero, Georg Raffelt, and Seokhoon Yun for useful discussions or remarks.
We also thank Steen Hannestad and Georg Raffelt for correspondence and for updating Ref.~\cite{Hannestad:2003yd}. 
This work was supported by IBS under the project code, IBS-R018-D1. 

\vspace{12pt}

\noindent \textit{Note added:} 
In the first version of this paper, which appeared on the arXiv on Dec. 31, 2024, we pointed out \textit{two} numerical inconsistencies in previous works devoted to astrophysical bounds on the ADD model~\cite{Hannestad:2003yd,Casse:2003pj}, which set the most stringent limits for $n < 4$.
We also provided corrected limits on ADD, by following the methodology of~\cite{Hannestad:2003yd} but using the correct numerical values for the decay widths of KK gravitons and the constraint on neutron stars excess heating, $L_{\mathrm{max}} \lesssim L_\odot \times 10^{-5}$.
This was confirmed in the recently updated (March 31, 2025) version of Ref.~\cite{Hannestad:2003yd}, which also fixed other numerical factors of order $O(1)$, and suggested several interesting ways to further improve these limits. 
Therefore, the discussion in appendix \ref{app:astro_bounds_ADD} is now superseded by the updated version of Ref.~\cite{Hannestad:2003yd}.
However, we would like to stress that our results shown in Tables \ref{tab:tab1} and \ref{tab:tab2} differ from Ref.~\cite{Hannestad:2003yd} because other numerical factors were also changed (not just values of KK lifetimes and $L_{\mathrm{max}}$). For our case, where $n=1$ ADD limit is relevant, the strongest limit comes from Ref.~\cite{Casse:2003pj}, where our limit, $M_5 > 3.6 \times 10^4\,\,\tev$, is in fact close to the value that can be inferred from the discussion of Ref.~\cite{Hannestad:2003yd}. 
Therefore, we expect our results shown in Fig. \ref{fig:Gen_Clock} and \ref{fig:Gen_Clock_compare} to be robust, and we leave more precise study of astrophysical bounds (especially the $\lambda^2 \gg 1$ regime of GLD, which can originate from heterotic string theory~\cite{Im:2018dum} and bears a resemblance to the dark dimension scenario~\cite{Montero:2022prj}) for future work.

\appendix
\section{Background solution for the GLD model}
\label{app:GLDBg}

The GLD action in the Einstein frame is given in Eq. (\ref{GLD_E}):
\dis{
{\cal S}=\int d^4x \int_{-\pi R}^{\pi R} dy \sqrt{-g}M_5^3\, &\left( \frac{1}{2} {\cal R} - \frac{1}{2} \partial^M S \partial_M S - V(S) -V^b(S)\right),
}
where
\bea
V(S) &=& -2k_b^2  e^{-2\lambda S/\sqrt{3}},\\  
V^b(S) &=& -\frac{e^{-\lambda S/\sqrt{3}}}{\sqrt{g_{55}}} \left[4k_0 \delta(y) +4k_\pi \delta(y-\pi R) \right].
\eea

Let us take an ansatz for the metric as
\dis{
ds^2 = e^{2\sigma_1(y)} \eta_{\mu \nu} dx^{\mu} dx^{\nu} + e^{2 \sigma_2(y)} dy^2.
}
This metric has to satisfy the Einstein equation.
The Einstein equation is
\dis{
G_{MN} = \frac{1}{M_5^3} T_{MN},
}
where
\bea
G_{MN} &=& {\cal R}_{MN} - \frac12 g_{MN} {\cal R}, \\
\frac{1}{M_5^3} T_{MN} &=&  \partial_M S \partial_N S - g_{MN} \left(\frac{1}{2} g^{AB} \partial_A S \partial_B S +V(S)+V^b(S) \right) + \delta_{M 5} \delta_{N 5} \, g_{MN} V^b(S). \nonumber \\
\eea
The $\mu\nu$-components of the Einstein equation gives
\dis{
3\left( \sigma_1''+2(\sigma_1')^2-\sigma_1'\sigma_2'\right)&= \\
- \frac12 (S')^2 + 2k_b^2 &e^{2(\sigma_2 - \lambda S/\sqrt{3})}
-e^{\sigma_2-\lambda S/\sqrt{3}} \left[-4k_0 \delta(y) - 4k_\pi \delta(y-\pi R) \right], \label{Emn}
}
while the $55$-component yields
\dis{
6(\sigma_1')^2 = \frac12 (S')^2 + 2k_b^2 e^{2(\sigma_2 -\lambda S/\sqrt{3})}. \label{E55}
}
Here the prime ($'$) denotes the derivative with respect to the coordinate $y$.
Combining Eq. (\ref{Emn}) and Eq. (\ref{E55}), we obtain
\bea
&&3 \sigma_1'' - 3\sigma_1' \sigma_2' + {S'}^2 = -e^{\sigma_2-\lambda S/\sqrt{3}} \left[-4k_0 \delta(y) - 4k_\pi \delta(y-\pi R) \right], \label{E1}\\ 
&&12{\sigma_1'}^{2} - {S'}^2 = 4k_b^2 e^{2(\sigma_2 - \lambda S/\sqrt{3})}. \label{E2}
\eea

On the other hand, the equation of motion for $S$ is
\dis{
 \partial_M (\sqrt{-g} g^{MN} \partial_N S) = \sqrt{-g} \frac{\partial }{\partial S} (V(S)+V^b(S)), 
}
which gives
\dis{
\sqrt{3} S'' + \sqrt{3} (4\sigma_1' - \sigma_2') S' = 4\lambda k_b^2  e^{2(\sigma_2 -\lambda S/\sqrt{3})}-\lambda e^{\sigma_2-\lambda S/\sqrt{3}} \left[-4k_0 \delta(y) - 4k_\pi \delta(y-\pi R) \right]. \label{Seq}
}

We find a solution to the equations of motion in Eq. (\ref{E1}), Eq. (\ref{E2}), and Eq. (\ref{Seq}) as
\bea
\sigma_1 &=& k_1 |y| + c_1, \\
\sigma_2 &=& k_2 |y| + c_2, \\
 \frac{\lambda}{\sqrt{3}}S &=&  \sigma_2 + \frac{\lambda}{\sqrt{3}}S_0, \\
 k_2 &=& \lambda^2 k_1 = \frac{2 \lambda^2 }{\sqrt{3(4-\lambda^2)}}k_b\, e^{-\lambda S_0/\sqrt{3}} , \\
 k_0 \,e^{-\lambda S_0/\sqrt{3}} &=& -k_\pi\, e^{-\lambda S_0/\sqrt{3}}  = \frac{3}{2} k_1,
\eea
where $c_1, c_2$, and $S_0$ are some constants. 
The constants $c_1$ and $c_2$ can be absorbed by a coordinate transformation as follows without loss of generality.
\dis{
x^\mu \rightarrow e^{-c_1} x^\mu, \quad y \rightarrow y e^{-c_2}.
}
On the other hand, the constant $S_0$ corresponds to the vacuum expectation value of $S$ at $y=0$, which may be determined by an additional brane potential \cite{Giudice:2017fmj}.
$S_0$ determines physical curvature scale of the extra dimension by
\dis{
k_b^{\rm phy} &= k_b e^{-\lambda S_0/\sqrt{3}}, \\
k_0^{\rm phy} &= k_0 e^{-\lambda S_0/\sqrt{3}}, \\
k_\pi^{\rm phy} &= k_\pi e^{-\lambda S_0/\sqrt{3}}. 
}

\section{Astrophysical bounds on large extra dimensions}
\label{app:astro_bounds_ADD}
The Lagrangian describing KK gravitons interactions with the SM is given by Eq. (\ref{lag}), which we write in the following form:
\be
  \mathcal{L} \supset -\frac{1}{\Lambda_n} h_{\mu \nu}^{(n)}\, T^{\mu\nu}_{\rm SM}\,
\ee
which allows to compute, \eg, the decay widths of KK gravitons. 
Since astrophysical bounds affect only light gravitons, the relevant decay channels are $\gamma\gamma$, $\nu\nu$, and $e^+ e^-$.
The analytical forms of the decay widths that were used in \cite{Hannestad:2003yd,Casse:2003pj} are given in \cite{Han:1998sg}.
The decay widths were also obtained in, \eg, Ref.~\cite{Lee:2013bua}, which confirmed these results up to a factor of $1/4$ for each decay width - the difference originates in the extra factor of $1/2$ in the Lagrangian of \cite{Han:1998sg} - compare Eq. 33 in \cite{Han:1998sg} to Eq. 4 in \cite{Lee:2013bua} - we follow the convention of \cite{Lee:2013bua}.

We note that there is an inconsistency in \textit{numerical} values for lifetimes of gravitons given by Eq. 47 and 50 in Ref.~\cite{Han:1998sg}.
Indeed, in the limit $m_G \gg m_V$, where $V$ indicates a massive gauge boson,
\be
    \tau_{G \to V V} \simeq \frac{5 \times 10^2}{\kappa^2 m_G^3} \simeq \frac12 \tau_{G \to \gamma\gamma} \simeq 30 \,\mathrm{yr} \left(\frac{100\,\gev}{m_G}\right)^3 = 3\times 10^{10} \,\mathrm{yr} \left(\frac{0.1\,\gev}{m_G}\right)^3\,,
\ee
which is larger than the value given by Eq. 47, $\tau_{G \to \gamma\gamma} \simeq 6\times 10^{9} \,\mathrm{yr} \left(\frac{0.1\,\gev}{m_G}\right)^3$ by a factor of 10.

Let us briefly recall the results of Ref.~\cite{Hannestad:2003yd}.
The rate at which a single graviton loses energy is determined by
\be
    Q_n = \frac{1}{\Lambda_n^2} \sigma_N n_B^2 T^{7/2}m_N^{-1/2} \,,
\ee
and to obtain the total rate, one needs to sum over all the $n\in \mathbb{N}$.
Since the mass splitting between consecutive gravitons is small (with exception of the RS in the $k\simeq M_5$ region), and since we consider models with $n=1$ extra dimensions, one can substitute the summation with the following integration measure:
\be
    \sum_n \to \, 2 R \int \left|\frac{dn}{dm}\right| dm\,,
\label{eq:density_ADD}
\ee
where $\left|\frac{dn}{dm}\right|=1/\left|\frac{dm}{dn}\right|$ can be easily computed from Eq. (\ref{KKm}).

In Table~\ref{tab:tab1}, we present the results of our analysis. We stress that i) we followed the methodology developed in~\cite{Hannestad:2001jv,Hannestad:2001xi,Hannestad:2003yd} to obtain the relevant fluxes, in particular Eq. 28, 47, and 55 in~\cite{Hannestad:2003yd}
and ii) we did not attempt to update the limits by using more modern data.
Regarding the first point, our improvements are due to \textit{numerical} factors, while the relevant analytical expressions in the mentioned papers, \eg,~\cite{Han:1998sg,Hannestad:2003yd,Casse:2003pj} are unchanged.
As for the latter point, we leave such task for a future work.

We believe that the difference between our limits and those given in~\cite{Hannestad:2003yd,Casse:2003pj} stems from the fact that both of these papers used Eq. 47 from~\cite{Han:1998sg}, which is too-small by a factor of $10$.\footnote{Note that for light KK gravitons, the total decay width satisfies $\tau=1/\Gamma=1/(\Gamma_{\gamma\gamma}+\Gamma_{3\nu\nu}+\Gamma_{e^+ e^-}) = 1/(2.25\Gamma_{\gamma\gamma})$, and the branching ratio into two photons is $\simeq 0.5$. However, Eq. 44 in~\cite{Hannestad:2003yd} directly follows Eq. 47 from~\cite{Han:1998sg}, and therefore the factor $\gamma$ in Eq. 46 in~\cite{Hannestad:2003yd} is too-large by a factor of $10$; we believe that Ref.~\cite{Casse:2003pj} used the same formula - see the equation without number described in the sentence ``and thus their lifetime...'' and Eq. 1 therein. Therefore, the photon flux given by Eq. 47, 55 in~\cite{Hannestad:2003yd} and Eq. 1 in~\cite{Casse:2003pj} was overestimated by 1 order of magnitude.}
Moreover, the NS excess heat bound on $\bar{M}_{4+n}$ stated in Table VI of Ref.~\cite{Hannestad:2003yd}, corresponds to $L_{\mathrm{max}} \lesssim L_\odot \times 5 \times 10^{-8}$, instead of the stated limit $L_{\mathrm{max}} \leq L_\odot \times 10^{-5}$ (factor of $\sim 200$ difference).
We were unable to trace the source of this discrepancy; we assume that the limit on $L_{\mathrm{max}}$ is $L_\odot \times 10^{-5}$, as written in Ref.~\cite{Hannestad:2003yd},\footnote{We also thank Georg Raffelt for correspondence on this point.} the paragraph below Eq. 56, since this is the result of~\cite{Larson:1998it}. 
Therefore, the effective photon flux relevant to the NS excess heat bound on $\bar{M}_{4+n}$ is weakened by a total factor of $\sim 10 \times 200 = 2000$ with respect to Ref.~\cite{Hannestad:2003yd}.
For an easy comparison with the tables from~\cite{Hannestad:2003yd}, in Table~\ref{tab:tab2}, we give both the upper limits on $R$ [m] and the lower limits on $\overline{M}_{4+n}$ [TeV]. 
Note that in the main body of our work, we consider $n=1$, \ie, $M_5=\overline{M}_{5}$ in the notation of~\cite{Hannestad:2003yd}; GB stands for Galactic bulge, and, for completeness, we also show the SNe limits from Ref.~\cite{Hannestad:2003yd}, which are unaffected.

\begin{table*}[h]
    \centering
    \hspace*{-1cm}
    \resizebox{\textwidth}{!}{%
    \begin{tabular}{|c||c|c|c|c|c|c|c|c|c|}
      \hline
      \hline
      $n$ & $1$ & $2$ & $3$ & $4$ & $5$ & $6$ & $7$ \\
      \hline
      \hline
      \thead{SN 1987A \\ (HR)} & \thead{($7.4\times 10^{2}$)} & \thead{($8.9$)} & \thead{($0.66$)} & \thead{($1.18\times 10^{-1}$)} & \thead{($3.5\times 10^{-2}$)} & \thead{($1.44\times 10^{-2}$)} & \thead{($7.2\times 10^{-3}$)} \\
      \hline
      \thead{EGRET \\SNe (HR)} & \thead{($3.4\times 10^{3}$)} & \thead{($28$)} & \thead{($1.65$)} & \thead{($2.54\times 10^{-1}$)} & \thead{($6.8\times 10^{-2}$)} & \thead{($2.56\times 10^{-2}$)} & \thead{($1.21\times 10^{-2}$)} \\
      \hline
      \hline
      \thead{NS excess \\ heat (HR)} & \thead{$1.2\times 10^{4}$ \\($1.61\times 10^{5}$)} & \thead{$100$ \\($7.01\times 10^{2}$)} & \thead{$5.3$ \\($25.5$)} & \thead{$0.75$ \\($2.77$)} & \thead{$0.2$ \\($0.57$)} & \thead{$0.065$ \\($0.17$)} & \thead{$2.84\times 10^{-2}$ \\($6.84\times 10^{-2}$)} \\
      \hline
      \thead{EGRET \\NS (HR)} & \thead{$1.4\times 10^{3}$ \\($2.93\times 10^{3}$)} & \thead{$22$ \\($38.6$)} & \thead{$1.7$ \\($2.65$)} & \thead{$0.3$ \\($0.43$)} & \thead{$0.084$ \\($0.116$)} & \thead{$3.3\times 10^{-2}$ \\($4.31\times 10^{-2}$)} & \thead{$1.53\times 10^{-2}$ \\($1.98\times 10^{-2}$)} \\
      \hline
      \thead{ EGRET \\ NS - GB\\ (CPBS)} & \thead{$3.6\times 10^{4}$ \\($7.8\times 10^{4}$)} & \thead{$2.5\times 10^{2}$ \\($4.5\times 10^{2}$)} & \thead{$12$ \\($19$)} & \thead{$1.5$ \\($2.2$)} & \thead{$3.4\times 10^{-1}$ \\($4.7\times 10^{-1}$)} & \thead{$1.1\times 10^{-1}$ \\($1.47\times 10^{-1}$)} & \thead{$4.6\times 10^{-2}$ \\($5.9\times 10^{-2}$)} \\
      \hline
      \hline
    \end{tabular}
    }
    \caption{
        Our limits on ADD parameter $\bar{M}_{4+n}$ [$\mathrm{TeV}$] obtained by following the discussion in~\cite{Hannestad:2003yd}. In parentheses, we indicate limits obtained by HR~\cite{Hannestad:2003yd} and CPBS~\cite{Casse:2003pj}.
    }
    \label{tab:tab1}
\end{table*}
\begin{table*}[h]
    \centering
    \hspace*{-1cm}
    \resizebox{\textwidth}{!}{%
    \begin{tabular}{|c||c|c|c|c|c|c|c|c|c|}
      \hline
      \hline
      $n$ & $1$ & $2$ & $3$ & $4$ & $5$ & $6$ & $7$ \\
      \hline
      \hline
      \thead{SN 1987A \\ (HR)} & \thead{($4.9\times 10^{2}$)} & \thead{($9.6\times 10^{-7}$)} & \thead{($1.14 \times 10^{-9}$)} & \thead{($3.82\times 10^{-11}$)} & \thead{($4.85\times 10^{-12}$)} & \thead{($1.21\times 10^{-12}$)} & \thead{($4.42\times 10^{-13}$)} \\
      \hline
      \thead{EGRET \\SNe (HR)} & \thead{($4.9$)} & \thead{($9.6\times 10^{-8}$)} & \thead{($2.47\times 10^{-10}$)} & \thead{($1.21\times 10^{-11}$)} & \thead{($1.93\times 10^{-12}$)} & \thead{($5.6\times 10^{-13}$)} & \thead{($2.29\times 10^{-13}$)} \\
      \hline
      \hline
      \thead{NS excess \\ heat (HR)} & \thead{$1.22\times 10^{-1}$ \\($4.44\times 10^{-5}$)} & \thead{$8.16\times 10^{-9}$ \\($1.55\times 10^{-10}$)} & \thead{$3.6\times 10^{-11}$ \\($2.58\times 10^{-12}$)} & \thead{$2.43 \times 10^{-12}$ \\($3.36 \times 10^{-13}$)} & \thead{$4.85 \times 10^{-13}$ \\($9.95 \times 10^{-14}$)} & \thead{$1.65 \times 10^{-13}$ \\($4.41 \times 10^{-14}$)} & \thead{$7.63\times 10^{-14}$ \\($2.46\times 10^{-14}$)} \\
      \hline
      \thead{EGRET \\NS (HR)} & \thead{$73.6$ \\($7.36$)} & \thead{$1.6\times 10^{-7}$ \\($5.13\times 10^{-8}$)} & \thead{$2.4\times 10^{-10}$ \\($1.12\times 10^{-10}$)} & \thead{$9.8\times 10^{-12}$ \\($5.46\times 10^{-12}$)} & \thead{$1.46\times 10^{-12}$ \\($9.13\times 10^{-13}$)} & \thead{$4.1\times 10^{-13}$ \\($2.8\times 10^{-13}$)} & \thead{$1.7\times 10^{-13}$ \\($1.21\times 10^{-13}$)} \\
      \hline
      \thead{ EGRET \\ NS - GB\\ (CPBS)} & \thead{$3.9\times 10^{-3}$ \\($3.9\times 10^{-4}$)} & \thead{$1.2\times 10^{-9}$ \\($3.8\times 10^{-10}$)} & \thead{$9.1\times 10^{-12}$ \\($4.2\times 10^{-12}$)} & \thead{$8.3\times 10^{-13}$ \\($4.7\times 10^{-13}$)} & \thead{$2\times 10^{-13}$ \\($1.3\times 10^{-13}$)} & \thead{$8\times 10^{-14}$ \\($5.4\times 10^{-14}$)} & \thead{$4.1\times 10^{-14}$ \\($2.9\times 10^{-14}$)} \\
      \hline
      \hline
    \end{tabular}
    }
    \caption{
        Same as Table~\ref{tab:tab1} but the limits are given on the parameter $R$ [$m$].
    }
    \label{tab:tab2}
\end{table*}

\section{Cross sections}
\label{app:xs}
\begin{figure}[h]
  \centering
  \includegraphics[scale=0.36]{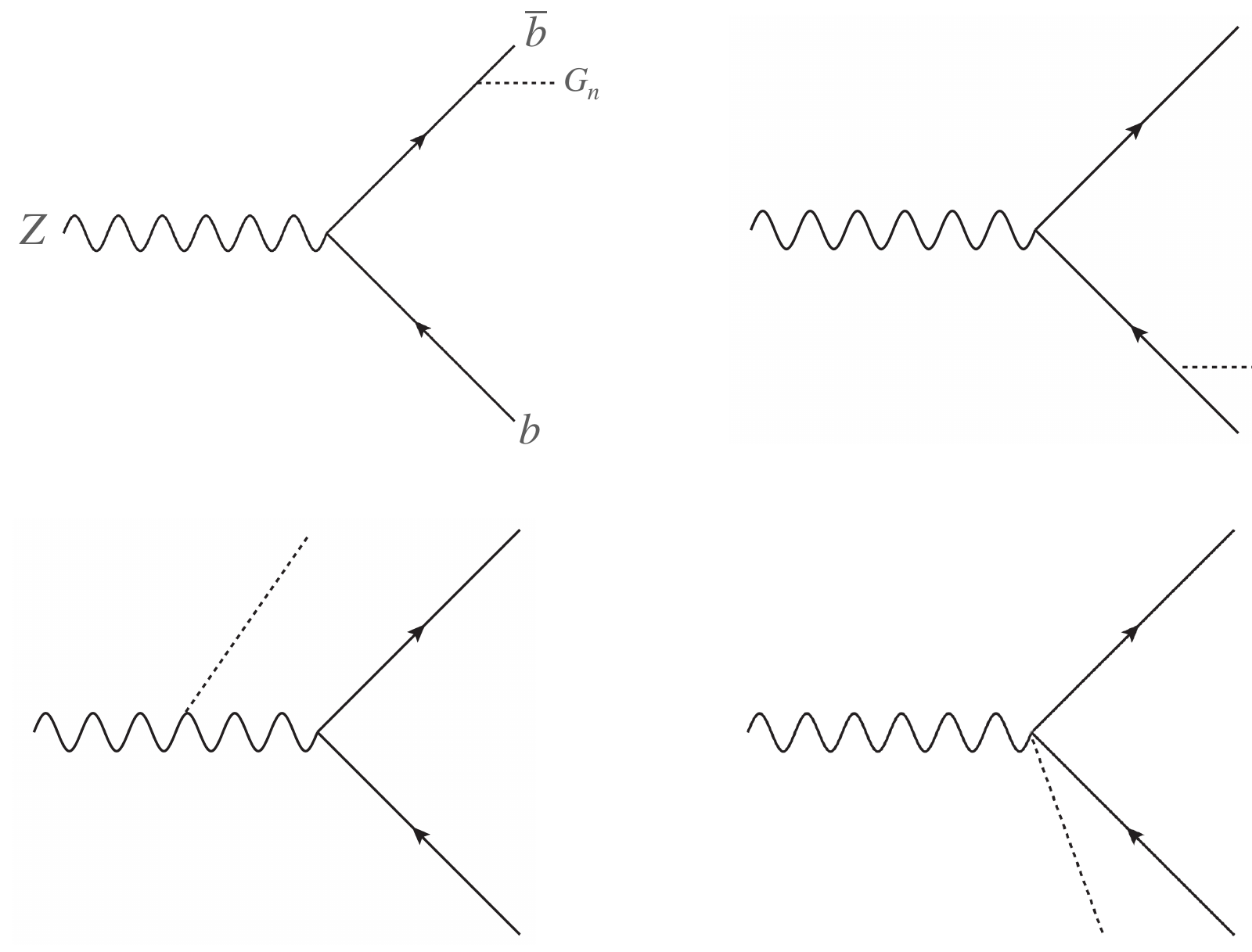}
  \caption{
    Feynman diagrams contributing to $Z\to f\bar{f}G_n$. The $e^+e^- \to G_n Z$ process is described by the same diagrams by using the crossing symmetry relations between the Mandelstam variables.
    }
    \label{fig:Z_qqbarGn}
\end{figure}

Below we provide the averaged amplitudes squared for the processes relevant to our analyses: $e^+e^- \to G_n \gamma$ or $e^+e^- \to G_n Z$ at FCC-ee and CLIC, and $Z\to f\bar{f}G_n$, where $f$ is charged SM fermion, at LLP mode of FCC-ee. The Feynman diagrams describing the latter process are shown in fig. \ref{fig:Z_qqbarGn}.
In order to obtain the corresponding cross sections, we follow the discussion in the kinematics section of the Particle Data Group \cite{ParticleDataGroup:2022pth}. In particular, $s$ and $t$ denote the Mandelstam variables.

\beq
\label{eq:amp2_eeGZ}
|M|_{e^+e^- \to G_n Z}^2 &= -\frac{e^2 \left(g_A^2+g_V^2\right)}{12 \Lambda^2 t^2 \left(\Gamma_Z^2 m_Z^2+\left(m_Z^2-s\right)^2\right) \left(m_{G_n}^2+m_Z^2-s-t\right)^2} \times \\
& \bigg[ 3 m_{G_n}^8 \left(m_Z^6-m_Z^4 (2 s+t)+m_Z^2 s   (s+2 t)-s t (s+4 t)\right) \nonumber\\
& +m_{G_n}^6 \bigg(6 m_Z^8-3 m_Z^6 (6 s+5 t)+m_Z^4 \left(18 s^2+57 s t+13   t^2\right) \nonumber\\
& -m_Z^2 \left(6 s^3+45 s^2 t+90 s t^2+4 t^3\right)+3 s t \left(s^2+11 s t+16   t^2\right)\bigg) \nonumber\\
& +m_{G_n}^4 \bigg(3 m_Z^{10}-3 m_Z^8 (4 s+5 t)+2 m_Z^6 \left(9 s^2+45 s t+16 t^2\right) \nonumber\\
& -2 m_Z^4 \left(6 s^3+57 s^2 t+110 s t^2+16 t^3\right)+3 m_Z^2 \left(s^4+14 s^3 t+64 s^2 t^2+76 s t^3+4 t^4\right) \nonumber\\
& -3 s t \left(s^3+12 s^2 t+36 s t^2+28 t^3\right)\bigg) \nonumber\\
& +m_{G_n}^2 t \bigg(-3 m_Z^{10}+m_Z^8 (33 s+13 t)-2 m_Z^6 \left(30 s^2+75 s t+16 t^2\right) \nonumber\\
& +m_Z^4 \left(36 s^3+206 s^2 t+276 s t^2+34 t^3\right)-3 m_Z^2 \left(3 s^4+34 s^3 t+96 s^2 t^2+76 s t^3+4 t^4\right) \nonumber\\
& +3 s \left(s^4+11 s^3 t+36 s^2 t^2+50 s t^3+24 t^4\right)\bigg) 
-4 t^2 \left(-m_Z^2+s+t\right)^2 \times \nonumber\\
& \bigg(m_Z^4 (5 s+t)-m_Z^2 t (7 s+t) +3 s \left(s^2+2 s t+2 t^2\right)\bigg) \bigg]\,, \nonumber
\eeq
\beq
\label{eq:amp2_ZGee}
|M|_{Z \to G_n e^+e^-}^2 &= -\frac{e^2 \left(g_A^2+g_V^2\right)}{12 \Lambda^2 s_{12}^2 \left(\Gamma_Z^2 m_Z^2+\left(m_Z^2-s_{23}\right)^2\right) \left(m_{G_n}^2+m_Z^2-s_{12}-s_{23}\right)^2} \times \\
& \bigg[ 3 m_{G_n}^8 \left(m_Z^6-m_Z^4 (s_{12}+2 s_{23})+m_Z^2 s_{23} (2 s_{12}+s_{23})-s_{12} s_{23} (4 s_{12}+s_{23})\right)  \nonumber \\
& +m_{G_n}^6 \bigg(6 m_Z^8-3 m_Z^6 (5 s_{12}+6 s_{23})+m_Z^4 \left(13 s_{12}^2+57 s_{12} s_{23}+18 s_{23}^2\right) \nonumber \\
& -m_Z^2 \left(4 s_{12}^3+90 s_{12}^2 s_{23}+45 s_{12} s_{23}^2+6 s_{23}^3\right)+3 s_{12} s_{23} \left(16 s_{12}^2+11 s_{12} s_{23}+s_{23}^2\right)\bigg) \nonumber \\
& +m_{G_n}^4 \bigg(3 m_Z^{10}-3 m_Z^8 (5 s_{12}+4 s_{23})+2 m_Z^6 \left(16 s_{12}^2+45 s_{12} s_{23}+9 s_{23}^2\right) \nonumber \\
& -2 m_Z^4 \left(16 s_{12}^3+110 s_{12}^2 s_{23}+57 s_{12} s_{23}^2+6 s_{23}^3\right) \nonumber \\
& +3 m_Z^2 \left(4 s_{12}^4+76 s_{12}^3 s_{23}+64 s_{12}^2 s_{23}^2+14 s_{12} s_{23}^3+s_{23}^4\right) \nonumber \\
& -3 s_{12} s_{23} \left(28 s_{12}^3+36 s_{12}^2 s_{23}+12 s_{12} s_{23}^2+s_{23}^3\right)\bigg)\nonumber \\
& +m_{G_n}^2 s_{12} \bigg(-3 m_Z^{10}+m_Z^8 (13 s_{12}+33 s_{23})-2 m_Z^6 \left(16 s_{12}^2+75 s_{12} s_{23}+30 s_{23}^2\right) \nonumber \\
& +m_Z^4 \left(34 s_{12}^3+276 s_{12}^2 s_{23}+206 s_{12} s_{23}^2+36 s_{23}^3\right)\nonumber \\
& -3 m_Z^2 (4 s_{12}^4+76 s_{12}^3 s_{23}+96 s_{12}^2 s_{23}^2 +34 s_{12} s_{23}^3+3 s_{23}^4) \nonumber \\
& +3 s_{23} \left(24 s_{12}^4+50 s_{12}^3 s_{23}+36 s_{12}^2 s_{23}^2+11 s_{12} s_{23}^3+s_{23}^4\right)\bigg)\nonumber \\
& -4 s_{12}^2 \left(-m_Z^2+s_{12}+s_{23}\right)^2 \nonumber\times \\
& \times \left(m_Z^4 (s_{12}+5 s_{23})-m_Z^2 s_{12} (s_{12}+7 s_{23}) +3 s_{23} \left(2 s_{12}^2+2 s_{12} s_{23}+s_{23}^2\right)\right) \bigg] \nonumber \,,
\eeq
\beq
\label{eq:amp2_eeG}
  |M|_{e^+e^- \to G_n}^2 &\simeq \frac{\pi s^2}{10 \Lambda^2} \,,
\eeq
where $G_n$ is a KK graviton with mass $m_{G_n}$ and universal coupling $\Lambda$, we neglected the masses of the electrons, and $g_V=1/(4 \sin\theta_W \cos\theta_W)$, $g_A=1/(4 \sin\theta_W \cos\theta_W)(1-4\sin^2\theta_W)$. 
Analogous formula holds for the $Z$ decays involving quarks, \eg, $Z\to G_n b\bar{b}$, where one replaces the fermion masses and couplings in trivial way.
The ranges of the integration variables, $s_{23}$ and $s_{12}$, are taken from the PDG.

\bibliography{bibliography}
\bibliographystyle{utphys}

\end{document}